\documentclass{IEEEtran}

\usepackage{times,amsmath,epsfig,multirow}
\usepackage{algorithmic,algorithm}
\usepackage{array}
\usepackage{color}
\usepackage{tablefootnote}
\usepackage{amssymb, mathrsfs, amsfonts}
\usepackage{graphicx}
\usepackage{url}
\usepackage{amsmath}
\usepackage{cite}
\usepackage[caption=false]{subfig}

\def\BibTeX{{\rm B\kern-.05em{\sc i\kern-.025em b}\kern-.08em
    T\kern-.1667em\lower.7ex\hbox{E}\kern-.125emX}}

\def\max{{\mathrm{max}}}

\def\si{{\mathsf{ref}}}

\def\TX{{\mathsf{TX}}}
\def\RX{{\mathsf{RX}}}
\def\BP{{\mathsf{BP}}}
\def\BE{{\mathsf{BE}}}

\def\gc{{\mathrm{gc}}}
\def\f{{\mathrm{f}}}

\begin{document}

\title{Cable Diagnostics with Power Line Modems for Smart Grid Monitoring}
\author{{Yinjia Huo},~\IEEEmembership{Student Member, IEEE},
{Gautham Prasad},~\IEEEmembership{Student Member, IEEE},
{Lazar Atanackovic}, 
{Lutz Lampe},~\IEEEmembership{Senior Member, IEEE}, 
and {Victor C.M. Leung},~\IEEEmembership{Fellow, IEEE} }

\maketitle

\begin{abstract}
Remote monitoring of {electrical cable conditions is an essential characteristic of the next-generation} smart grid, which features the ability to consistently surveil and control the grid infrastructure. In this paper, we propose a technique that harnesses power line modems (PLMs) as sensors for monitoring cable health. We envisage that all or most of these PLMs have already been deployed for data communication purposes {and focus} on the distribution grid or neighborhood area networks in the smart grid. For such a setting, we propose a machine learning (ML) based framework for automatic {cable diagnostics} by continuously monitoring the cable status to identify, assess, and locate possible degradations. As part of our technique, we also synthesize  state-of-the-art reflectometry methods within the PLMs to extract beneficial features for effective performance {of our proposed ML solution}.  
Simulation results demonstrate the effectiveness of our solution under {different aging conditions and varying load configurations.} Finally, we reflect on our proposed diagnostics method by evaluating its robustness and comparing it with existing alternatives.
\end{abstract}


\IEEEpeerreviewmaketitle

\section{Introduction}
\label{sec:introduction}
A salient feature of smart grids is ubiquitous monitoring and control targeted for purposes such as frequency regulation, demand response, asset management, and anomaly detection~\cite{farhangi2010path}. As a smart grid monitoring technique, cable health monitoring falls into the latter two categories, since underground cables are crucial assets of the grid infrastructure and are widely deployed in both transmission and distribution systems due to their aesthetics, robustness to weather conditions, and reduced impact on the environment~\cite{fairley2018utility, orton2013history}. Cable in-service failures lead to severe anomalies such as power outages. An ideal solution to avoid in-service failures in operating cables is to identify potential issues and anticipate faults in a non-destructive manner~\cite[Ch.~6]{gill2008electrical}. Such a proactive diagnostics approach, however, is known to be challenging as opposed to a reactive solution like hard fault detection~\cite{cozza2016echo}. Despite this, various diagnostics schemes have been developed in the past that attempt to identify and locate cable degradations~\cite[Ch. 6]{gill2008electrical},~\cite[Ch. 4]{andersen2012smart}.

A major drawback of the existing cable diagnostics schemes is the requirement of dedicated equipments to conduct the tests~\cite[Ch. 6]{gill2008electrical},~\cite{dubickas2006line}. To counter this, we propose reusing power line modems (PLMs) that are installed in the grid for smart-grid communications purposes~\cite{galli2011grid, mengi2017itut}. Any two PLMs in the network communicating with each other, regularly estimate the power line communication (PLC) channel condition between them for efficient end-to-end communications. We have shown in our recent works that this estimated channel condition also provides insight into the health of the power cables~\cite{forstel2017grid, huo2018cable, huo2018grid}. PLMs operating with broadband power line communication (BB-PLC) technology use analog-to-digital converters functioning at sampling rates of up to $200~\mathrm{MHz}$~\cite{yonge2013overview}. Signals that are sampled at such high rates provide distinctive information about the deterioration of cable insulation dielectric properties as the cables age and degrade~\cite{griffiths2006invisible},~\cite[Ch. 2, Ch. 7]{Neier2015cable},~\cite{Abou-Dakka:2008aa, werelius2001development}. This means that changes in the cable dielectric properties manifest themselves as variations in the BB-PLC channels estimated inside legacy PLMs. This enables us to design a PLM-based proactive diagnostics technique to identify potential issues and anticipate faults along the cable by monitoring the communication channel conditions. Additionally, this scheme can also operate online, unlike {several other conventional methods} that require cables to be de-energized during testing~\cite[Ch. 6]{gill2008electrical}.

Another major disadvantage of conventional diagnostics solutions is that most of them require technical staff with expertise to analyze the measured signals (current and/or voltages) and determine the cable health status~\cite[Ch. 6]{gill2008electrical}. This manual interpretation introduces added costs and slows down the diagnostics procedure. Furthermore, varying network load conditions result in signal changes that are on occasions similar to those caused due to cable degradations. This renders manual diagnosis susceptible to missed detections and false alarms {(FAs)}. To overcome these drawbacks, we propose augmenting our PLM-based approach with machine learning (ML) techniques to intelligently analyze the communication channel conditions estimated by the PLMs, and design a self-reliant and an automated cable diagnostics solution that can be implemented remotely.


\subsection{Contributions}
In this paper, we present for the first time, an ML framework for automated cable diagnostics that builds on PLM signal measurements and applies a multi-step cooperative scheme to progressively detect, assess, and locate possible cable degradations. We describe the underlying signal propagation principles, the extraction of suitable signal features that are indicative of cable degradations, the choice of appropriate ML algorithms to infer cable health using those features, and their training through synthetically generated data.

The performance of the ML algorithm relies on the quality of the chosen features. {Inspired by state-of-the-art joint-time-frequency domain reflectometry (JTFDR) solutions, which are fairly effective in cable anomaly detection and localization~\cite{wang2010application}, we design features extracted from the JTFDR waveform, together with other features extracted from the estimated communication channel.} However, adopting conventional JTFDR requires bulky expensive external devices and experienced technicians to be dispatched onsite, which are not consistent with the target characteristics of our solution that we previously outlined. To overcome this overhead, we devise a method to synthesize JTFDR within the PLMs, called PLM-JTFDR. Thereby, we incorporate PLM-JTFDR as part of our ML diagnostics solution by extracting features from its resultant waveforms.

For performance evaluations of our solution, we adopt a cable aging model from the literature~\cite{crine2005water, mugala2005high}, refine it for faithful emulation of realistic cable degradations, and further apply it to generate the PLC channels using a bottom-up approach. We generate channels under different network loads and cable degradation conditions for extracting features to train and test our machines. We use a sufficient number of samples while training a machine such that the ML algorithm performance is saturated. We note that since practical operation of our solution involves off-site training, we are not limited by the number of training samples used. For practical implementation, several machines are trained for a variety of diagnostics tasks and are loaded onto the PLMs in the grid through software/firmware upgrades in a remote manner.

In order to verify the effectiveness of our schemes, we present a comprehensive simulation-based evaluation of our proposed solution under a generic network topology with varying load conditions and arbitrary cable aging profiles. Further, to judge the effectiveness of our proposed solution under real-world scenarios, we critically evaluate its robustness by emulating non-idealities that could be encountered during practical deployment. In particular, we conduct a robustness and sensitivity analysis of our solution {with behavior variations from the adopted degradation models.}

%
%


\subsection{Outline}

The rest of the paper is organized as follows. In Section~\ref{sec:ageingModel}, we introduce the modeling of cable degradation, including its growth and the dielectric property change it causes within the insulation. We then describe the {modeling and generation} of PLC channels subject to a particular degradation condition under a considered network topology in Section~\ref{sec:plcModel}. In Section~\ref{sec:ML}, we introduce our proposed ML framework for automated cable {diagnostics}, including PLM-JTFDR for feature {extraction}. We present the simulation results in Section~\ref{sec:numerical}, and discuss the potential challenges for our proposed solution and compare it with prior works in Section~\ref{sec:discussion}. Conclusions are drawn in Section~\ref{sec:conclusion}. {The specifics of PLM-JTFDR and the details of our ML implementation} are relegated to the Appendices~\ref{sec:plcJTFDR}-\ref{sec:avoid_overfitting}.

\textit{Nomenclature}: Throughout this paper, we use $\Re(x)$ to indicate the real part of a complex number $x$. We denote a uniform random distribution between $a$ and $b$ as $\mathscr{U}(a,b)$.

\section{Cable Aging Model} 
\label{sec:ageingModel}
Based on the insulation material used, power cables are classified into laminated and extruded types~\cite[Ch. 6.2.3]{gill2008electrical}. While laminated cables suffer mainly from thermal degradations, extruded ones are mostly susceptible to electrical aging, i.e., developing water-treeing (WT) and electrical-treeing~\cite{densley2001ageing}. In this paper, we focus on investigating cable aging caused by WT degradations, since WT is a prominent cause {for premature failures of extruded cables}~\cite{patsch1999water},~\cite[Ch. 6]{gill2008electrical}. In addition, focusing on WT degradations has the added benefit that many works in the literature, e.g.,~\cite{crine2005water, mugala2005high} have developed a deterministic model for WT growth and thoroughly investigated the dielectric properties of WT-degraded extruded cross-linked polyethylene (XLPE) insulation, which is widely deployed~\cite{orton2013history, fairley2018utility, hanley2003general}.

\subsection{Cable Aging Profile}
\label{subsec:ageingProfile}

\begin{figure}[!t]
	\centering
	\includegraphics[width=7cm]{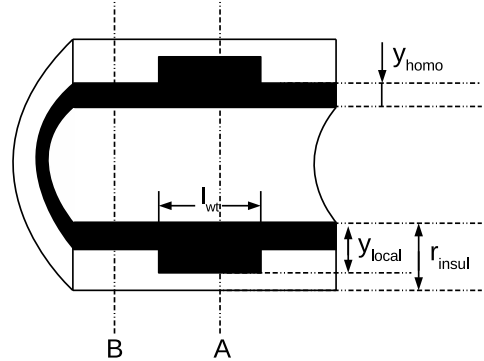}
	\caption{Longitudinal section of a cable insulation illustrating the cable aging profile. The shaded region indicates the portion of the insulation subjected to WT degradation while the unshaded region consists of the intact sections of the insulation and the {conductor-insulation interface}.}
\label{fig:ageingProfile}
\end{figure}


Investigations into the nature of WT in cable insulations have shown that under normal operating conditions, power cables develop near-uniform WT degradations across their length as they age~\cite{crine2005water, werelius2001development}.
However, water ingress or local defects may lead to salient localized WT degradations~\cite{crine2005water}. Therefore, for faithful emulation, we model the aging condition along the cable with an aging profile as shown in Fig.~\ref{fig:ageingProfile}. The cable section develops a homogeneous WT degradation along its length with thickness $y_\textsf{homo}$, within the total insulation thickness of $r_\textsf{insul}$. In addition, a section of the cable, of length $\ell_{\mathsf{WT}}$, may also be subject to a localized WT degradation of thickness $y_{\mathsf{local}}$.

The growth of $y_\textsf{homo}$ can be related to the elapsed service time, {$t_\mathsf{sr}$}, as~\cite{crine2005water}
\begin{equation}
\label{eqn:CableDeg}
y_\mathsf{homo}= \left( {\frac{\alpha_0 {\nu_0} f_0 F^2 \epsilon_0  \Re\{\epsilon_{w}\}  {{t_\mathsf{sr}}^{\frac{3}{2}}} }{Y}  } \right)^{\frac{1}{3}},
\end{equation}
where $F$ is the operating electric field strength (see Section~\ref{sec:plcModel}). The remaining parameters are all listed in Table~\ref{table:parameters} along with their values that we use in our model\footnote{Note that the properties of XLPE insulation material vary with the manufacturing process and depend on various factors including its density and crystallinity~\cite{tamboli2004crosslinked}. The parameters listed in Table~\ref{table:parameters} are values obtained under typical situations and serve as nominal parameters.}. Thus, an estimated value of $y_\mathsf{homo}$ can be used to obtain an \textit{equivalent age}, $t_\mathsf{eq}$, of a degraded cable, which can be computed as
\begin{equation}
\label{eqn:CableAge}
{t_\mathsf{eq}} = \sqrt[3]{\left( \frac{ Y \cdot y_\mathsf{homo}^3  } { \alpha_0 {\nu_0} f_0 F^2 \epsilon_0\Re\{\epsilon_{w}\} } \right)^ {2}}.
\end{equation}
This equivalent age provides an intuitive indication into the WT degradation severity that the cable has experienced.

\begin{table}
	\centering
	\caption{Cable Ageing Model Parameters~\cite{pruitt2005conventional,crine2005water, mugala2005high},~\cite[p. 794]{titow2012pvc}} \label{table:parameters}
	\begin{tabular}{c|c}
		\hline
		\hline
		Diffusion constant of water into the dielectric ($\alpha_0$) & $1.44 \times 10^4$ \tabularnewline
		Size of the free-volume voids {($\nu_0$)} & $2.5 \times 10 ^ {-28}{~\mathrm{m}^3}$ \tabularnewline
		Mains frequency on the line ($f_0$) & $60~\mathrm{Hz}$ \tabularnewline
		Absolute permittivity ($\epsilon_0$) & $8.8 \times 10^{-12}~\mathrm{F}/\mathrm{m}$ \tabularnewline
		Relative permittivity of non-degraded XLPE ($\epsilon_\mathsf{PE}$) & $2.3 - 0.001j$ \tabularnewline
		Dielectric {mechanical} yield strength ($Y$) & $2 \times 10^7~\mathrm{Pa}$ \tabularnewline
		Depolarization factor ($D$) & $\frac{1}{12}$ \tabularnewline
		Absolute water content in the WT region ($q_w$) & $0.06$ \tabularnewline
		Conductivity of water ($\sigma_w$) & $0.22~\mathrm{S}$ \tabularnewline
		\hline
		\hline
	\end{tabular} 
\end{table}

\subsection{Dielectric Properties of WT Degraded Cables}
\label{subsec:WTDielectric}
{The dielectric properties of WT degraded XLPE insulation material have been investigated in the literature, e.g., in~\cite{mugala2005high, stucki1993dielectric}.} 
The relative permittivity of WT degraded XLPE insulation (shown as the shaded region in Fig.~\ref{fig:ageingProfile}) can be computed as~\cite[Eq. 1]{stucki1993dielectric}
\begin{equation}
\label{eqn:WTRegion}
\epsilon_\mathsf{WT}= \epsilon_\mathsf{PE}\left(1+\frac{q_w(\epsilon_{w}-\epsilon_\mathsf{PE})}{D(1-q_w)(\epsilon_{w}-\epsilon_\mathsf{PE})}\right),
\end{equation}
where the relative permittivity of water is given by $\epsilon_w= 81-j\frac{\sigma_w}{2 \pi f \epsilon_0}$, for any operating frequency $f$. Please refer to Table~\ref{table:parameters} for descriptions and values of all other parameters. For the overall cross section of XLPE insulation with both WT degraded and intact regions, the equivalent relative permittivity can be computed using the series dielectrics model as~\cite[Eq. 6.3]{mugala2005high},
\begin{equation}
\label{eqn:EpsilonTotal}
\epsilon_\mathsf{total}=\left({\frac{y}{r_\mathsf{insul}}\frac{1}{\epsilon_\mathsf{WT}}+\frac{r_\mathsf{insul}-y}{r_\mathsf{insul}}\frac{1}{\epsilon_\mathsf{PE}}}\right)^{-1},
\end{equation}
where $y$ is the WT degradation depth. 

To further determine and establish degradation severities, we define $\gamma \triangleq \frac{y}{r_\mathsf{insul}}$ as the relative depth of the XLPE insulation that is impacted with the WT degradation. For a cross section with localized WT degradation ({e.g., cross section A} in Fig.~\ref{fig:ageingProfile}), we have $y=y_\textsf{local}$ and $\gamma_\mathsf{local}=\frac{y_\textsf{local}}{r_\mathsf{insul}}$. For other cross sections without localized WT degradation ({e.g., cross section B} in Fig.~\ref{fig:ageingProfile}), we have $y=y_\textsf{homo}$ and $\gamma_\mathsf{homo}=\frac{y_\textsf{homo}}{r_\mathsf{insul}}$. For an aging profile without any localized WT degradation, we have $\gamma=\gamma_\mathsf{homo}=\frac{y_\textsf{homo}}{r_\mathsf{insul}}$ along the entire length of the cable segment.

\section{PLC Channel Model}
\label{sec:plcModel}
Our next step involves generating several power line channels for healthy and degraded cables that will be used as training and testing samples in our ML framework. PLC channel modeling is typically addressed either through the top-down or the bottom-up approach. While the top-down approach could be computationally simpler, we adopt the bottom-up approach of modeling PLC channels in this investigation in order to accurately capture the influence of the insulation dielectric property changes resulting from a specific aging profile~\cite{versolatto2011mtl}.

For a given aging profile of cable insulation, each segment along the cable with the same degradation severity can be viewed as a uniform line with electrically small cross-sectional dimensions, in which condition PLC signals are transmitted in the quasi-transverse-electromagnetic (quasi-TEM) propagation mode, and the PLC channel can be modeled with the multi-conductor transmission line (MTL) theory~\cite[Ch. 1]{paul2008analysis}. We then concatenate each of these sections to obtain the overall PLC channel modeling. By using the WT degradation model of~\eqref{eqn:EpsilonTotal} in the MTL {per-unit-length (PUL)} parameter computations~{\cite[Eq. 4-11]{huo2018grid}}, we model the PLC channel for cable sections with arbitrary degradation conditions. For the software implementation, we feed the computed PUL parameters to an open source {bottom-up} PLC channel emulator of~\cite{gruber2015plc} to generate PLC channels.

\subsection{Network Topology and Loads}
\label{subsec:netTopo}

\begin{figure}[!t]
	\centering
	\includegraphics[width=7cm]{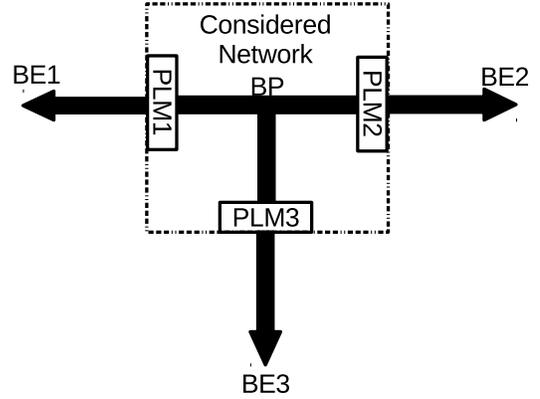}
	\caption{Portion of the distribution network considered.}
	\label{fig:networkTopology}
\end{figure}

Along with the knowledge of cable PUL parameters, channel generation also requires information about the network topology and the connected loads. Typical medium voltage (MV) distribution networks consist of few intermediate branches and can be divided into smaller $\textit{T}$ and chain networks~\cite{abeysinghe2018topological}.
Therefore, for a realistic grid emulation, we consider a $\textit{T}$-network with three PLMs, as shown in Fig.~\ref{fig:networkTopology}, with possible network extensions beyond each of the PLMs. {Each of the six branches in the considered network topology assumes an aging profile shown in Fig.~\ref{fig:ageingProfile} with the same $\gamma_\mathsf{homo}$.}
Without loss of generality, we consider a symmetrical topology with an equal separation distance of $500~\mathrm{m}$ between each PLM and the branch point (BP){~\cite{devoloModem}}. Further, we also place a $500~\mathrm{m}$ transmission line between each PLM and any further branch extensions (BEs). To emulate a range of realistic network extensions~\cite[Table 1.1]{berger2014mimo}, we consider equivalent load impedances randomly chosen in $\mathscr{U}(0,50)+j\cdot\mathscr{U}(-50,50){~\mathrm{\Omega}}$ between each pair of conductors {as shown in~\cite[Fig. 3]{huo2018cable}.}

\subsection{Cable Settings}
\label{subsec:wireConfig}
The final aspect of characterizing the PLC channel is to specify {the physical cable parameters}. In our evaluations, we consider the XLPE multi-core cable \textsf{N2XSEY}~\cite{HELUKABEL}, with equidistant conductor separations of {$d_\mathsf{cond}=15.88~\mathrm{mm}$, equal conductor radii {$r_\mathsf{cond}=3.99~\mathrm{mm}$} and a maximum rated voltage $V_0=12~\mathrm{kV}$.}
We then apply an approximate cylindrical geometry to compute the electric field {under $V_0$} at a distance $r_\mathsf{cond}$ from the center of the conductor~\cite[Eq. 1]{shu2013water}, where the electric field strength is at its maximum and is most prone for WT inception~\cite{shu2013water}, as
\begin{equation}
\label{eqn:FExpression}
F_\max=\frac {V_0} {r_\mathsf{cond} \ln \left(\frac{d_\mathsf{cond}}{2r_\mathsf{cond}}\right)}.
\end{equation}
{Considering} an expected maximum service time of $t_\max=30~\mathrm{years}${~\cite[Ch. 6]{gill2008electrical}} in~\eqref{eqn:CableDeg}, we derive the maximum homogeneous degradation severity, $\max(\gamma_\mathsf{homo})=0.0481$ under nominal conditions. Therefore, in our evaluations, we confine $\gamma_\mathsf{homo} \leq 0.05$, and to clearly distinguish a salient localized WT degradation, we let $\gamma_\mathsf{local} \geq 0.1$.

\subsection{PLC Channel Generation}
\label{subsec:generateCFR}
We use the PLC channel emulator of~\cite{gruber2015plc} to generate PLC channels in the frequency range of $2-30~\mathrm{MHz}$ with a frequency resolution of $\Delta f=24.414~\mathrm{kHz}$. 
The channel generator of~\cite{gruber2015plc} provides the end-to-end channel frequency response (CFR), $H_\f$, and the line access impedance, $\textbf{Z}_\textsf{in}$. Using $\textbf{Z}_\textsf{in}$ and the known PLM output impedance, we compute the reflection channel transfer function, $H_\mathsf{ref}$, which characterizes the behavior of the portion of the transmitted PLC signal that is {reflected into the modem~\cite{prasad2016enhancing}.} We therefore use both $H_\f$ and $H_\mathsf{ref}$ as training and testing samples for our ML framework. Note that both these channels are inherently estimated within the PLMs in some form, and hence introduces no additional overhead~\cite{yonge2013overview, prasad2016band}. While $H_\f$ is directly estimated within all legacy PLMs, $\textbf{Z}_\textsf{in}$ or $H_\mathsf{ref}$ can be obtained in either of two ways. Newer PLMs that support adaptive power boosting, regularly estimate $\textbf{Z}_\textsf{in}$ to compute the input return loss~\cite{yonge2013overview}. Alternatively, $H_\mathsf{ref}$ can also be acquired using in-band full-duplex (IBFD) communication functionality which inherently involves estimating the self-interference (SI) channel to cancel the echo~\cite{prasad2016band}. In the context of PLC, the echo/SI channel essentially indicates $H_\mathsf{ref}$~\cite{prasad2016band, prasad2017digitally}. In this work, we consider $H_\f$ and $H_\mathsf{ref}$ to be perfectly estimated by the PLMs in order to focus specifically on determining the potential of using PLC channel variations to estimate cable degradations.

\section{ML Framework for Cable Health Monitoring}\label{sec:ML}

In this section, we introduce our proposed ML solution for automated {cable diagnostics}.

\begin{figure}[!t]
	\centering
	\includegraphics[width=8cm]{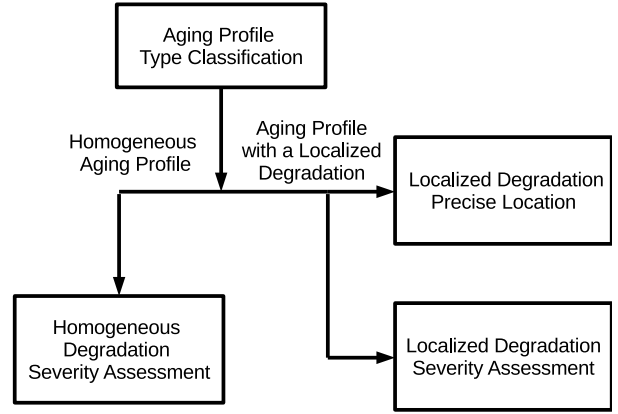}
	\caption{Our proposed multi-step cable diagnostics.}
	\label{fig:inference_procedure}
\end{figure}

\subsection{Cable Health Monitoring}
\label{subsec:multi_step}

We conduct the evaluation of cable health conditions through a series of procedures as shown in Fig.~\ref{fig:inference_procedure}. In the first stage, we perform an aging profile type classification, where we either detect a localized degradation (LD) or a homogeneous aging profile. When the cable has undergone homogeneous degradation, we assess its degradation severity. The case of a healthy cable is encompassed within this condition when a severity assessment results in zero degradation. In cases where the cable is subject to an {LD}, we perform a localization of this defect as well as an LD severity assessment.

\subsection{PLM-JTFDR}
\label{subsec:PLM_JTFDR}
The state-of-the-art in reflectometry for LD diagnostics is JTFDR, which transmits a reference signal that is customized based on the application, and then conducts advanced post-processing of the received reflected signal~\cite{wang2010application}. While JTFDR shows promising results in cable anomaly detection and localization, applying conventional JTFDR in its native form requires expensive external equipments as described in Section~\ref{sec:introduction}. To overcome the implementation drawbacks, we propose a new method called PLM-JTFDR, to synthesize the results of JTFDR operation, {i.e., to obtain the resultant JTFDR waveform, $h_\mathsf{JTFDR}$}, within PLMs by using $H_\mathsf{ref}$. We describe PLM-JTFDR in greater detail in Appendix~\ref{sec:plcJTFDR}.

Since JTFDR is limited to LD diagnostics due to its nature of operation, we consider {PLM-JTFDR} for developing parts of our solution focused on detecting, locating, and {assessing LDs}. An LD causes discontinuities in dielectric properties of cable insulations, which results in {parts of the PLC signal to be reflected back to the PLM from the locations of these discontinuities.} By determining the presence and locations of such reflections, which can be noticed as peaks {in the PLM-JTFDR waveforms}, an LD can be identified and localized. 

As an example, we study the use of {PLM-JTFDR} for localized cable diagnostics under a network topology shown in Fig.~\ref{fig:networkTopology}.
\begin{figure}
     \centering
     	\subfloat[]{\includegraphics[width=4cm]{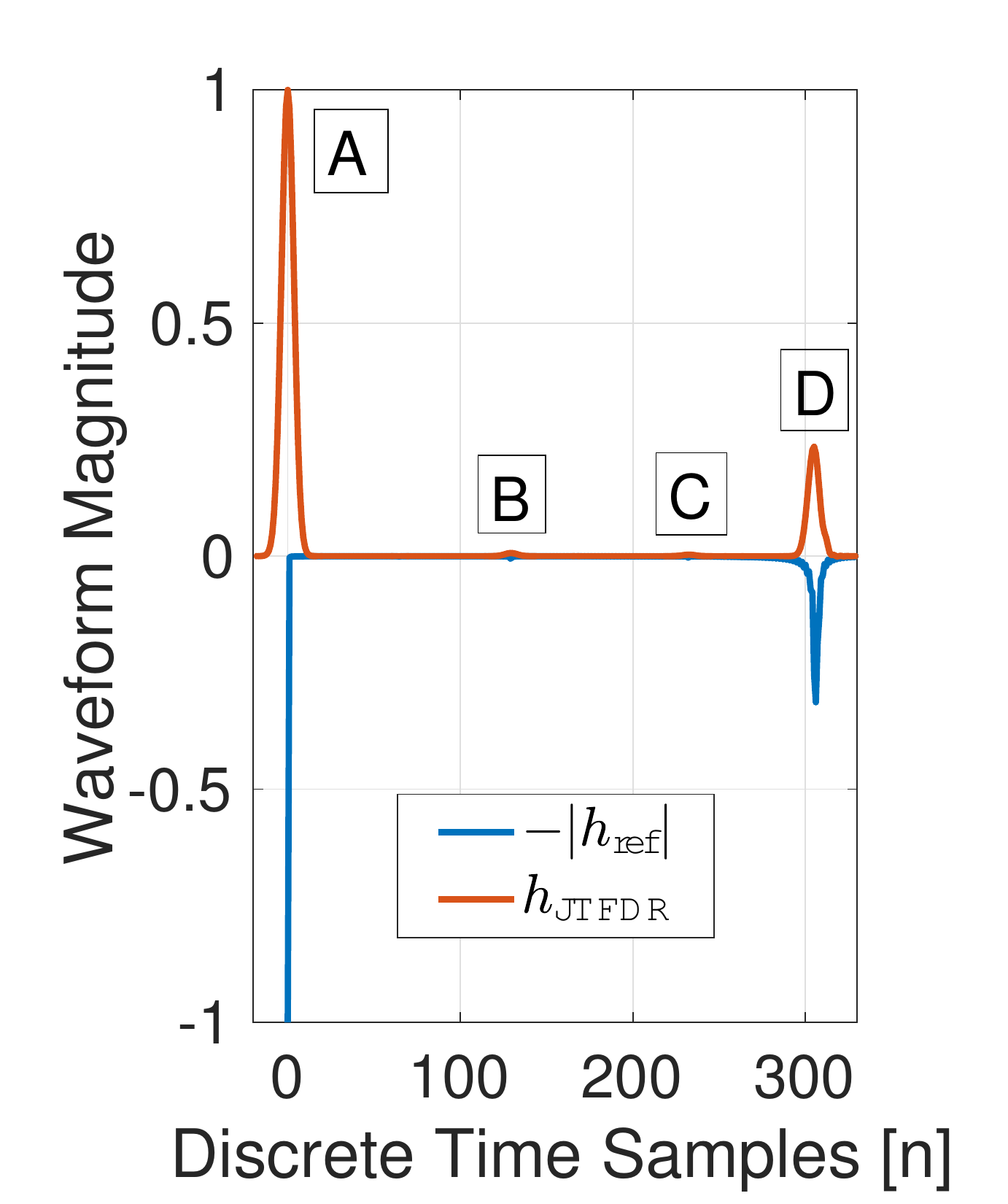} \label{fig:TX_Result_hsi_a}}
     	\subfloat[]{\includegraphics[width=4cm]{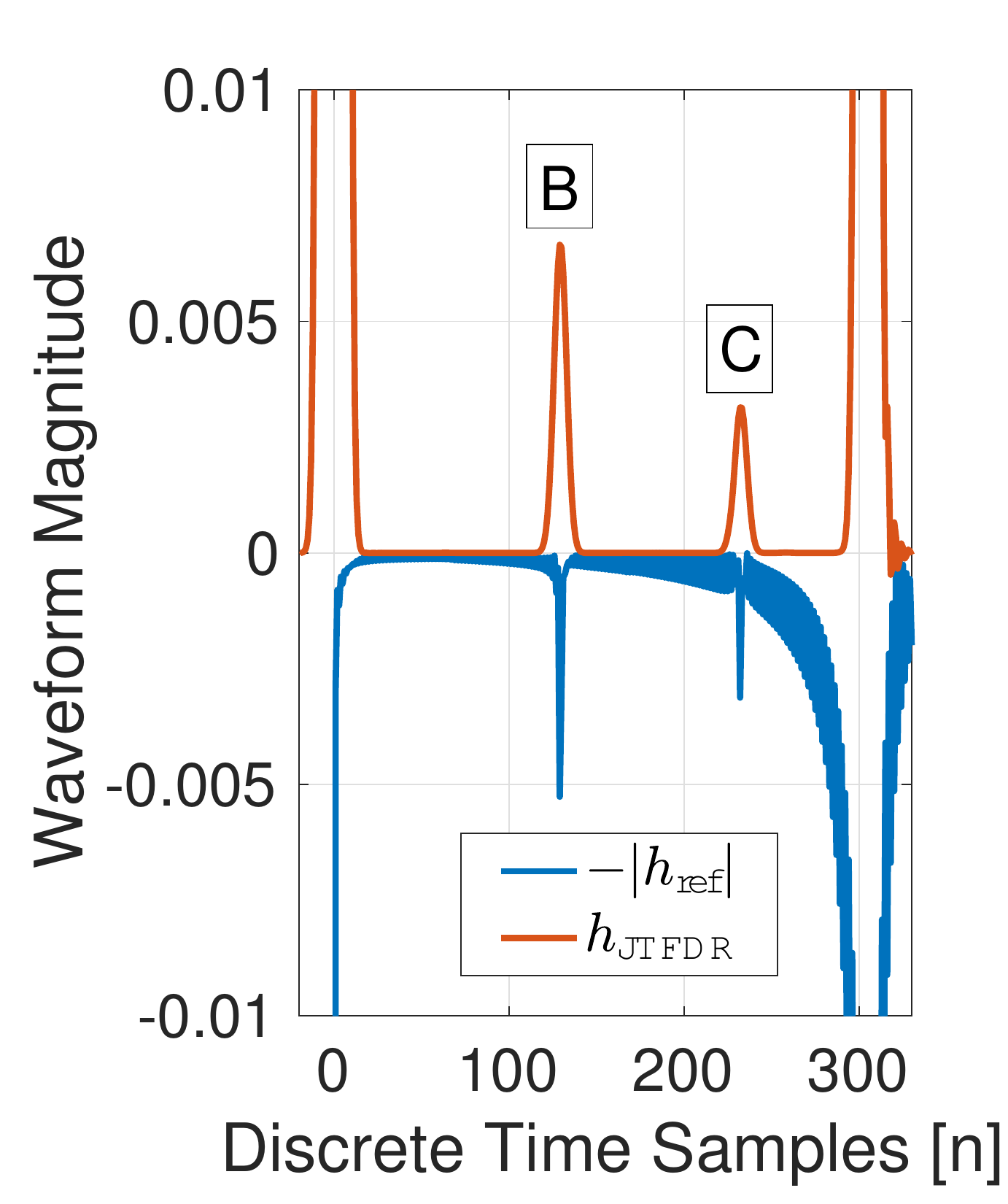} \label{fig:TX_Result_hsi_b}}
    \caption{(a) $h_\mathsf{JTFDR}$ and $-|h_\mathsf{ref}|$ {(normalized to the magnitude of peak A)} when a degradation is present between $\TX$ and $\BP$. (b) Zoomed at the degradation locations $B$ and $C$.}
	 \label{fig:TX_Result_hsi}
\end{figure}
We first run PLM-JTFDR with an aging profile that contains an LD located between $\mathsf{PLM1}$ and $\mathsf{BP}$. We set the LD to be of the minimum degradation severity, $\gamma_\mathsf{local}=0.1$, to evaluate the sensitivity of our methods to mild degradations. 
{We set an arbitrary degradation length of $166~\mathrm{m}$ to be present between $211~\mathrm{m}$ and $377~\mathrm{m}$ from $\mathsf{PLM1}$.} The resultant waveforms obtained for this setup is shown in Fig.~\ref{fig:TX_Result_hsi}\footnote{For illustration clarity, we only show the first 330 time samples of the signal, which contains all pertinent reflection peaks. The signal is normalized with respect to the magnitude of its first peak.}. For comparison, we also show $-|h_\mathsf{ref}|$ alongside, which is essentially the resultant waveform of conventional {time-domain reflectometry (TDR)}. We notice from Fig.~\ref{fig:TX_Result_hsi} that $h_\mathsf{JTFDR}$ is smoother and its peaks are more prominent than for $h_\textsf{ref}$. Therefore, compared to TDR, i.e., directly using $h_\textsf{ref}$ for cable diagnostics, PLM-JTFDR has superior sensitivity and performance to detect less salient localized WT degradations. The superior performance of our solution is more prominent when the LD is further away from the measurement point ($\mathsf{PLM1}$), e.g., between $\mathsf{PLM2}$ and $\BP$, as shown by the results in Fig.~\ref{fig:RX_Result_hsi}.

\begin{figure}
     \centering
     	\subfloat[]{\includegraphics[width=4cm]{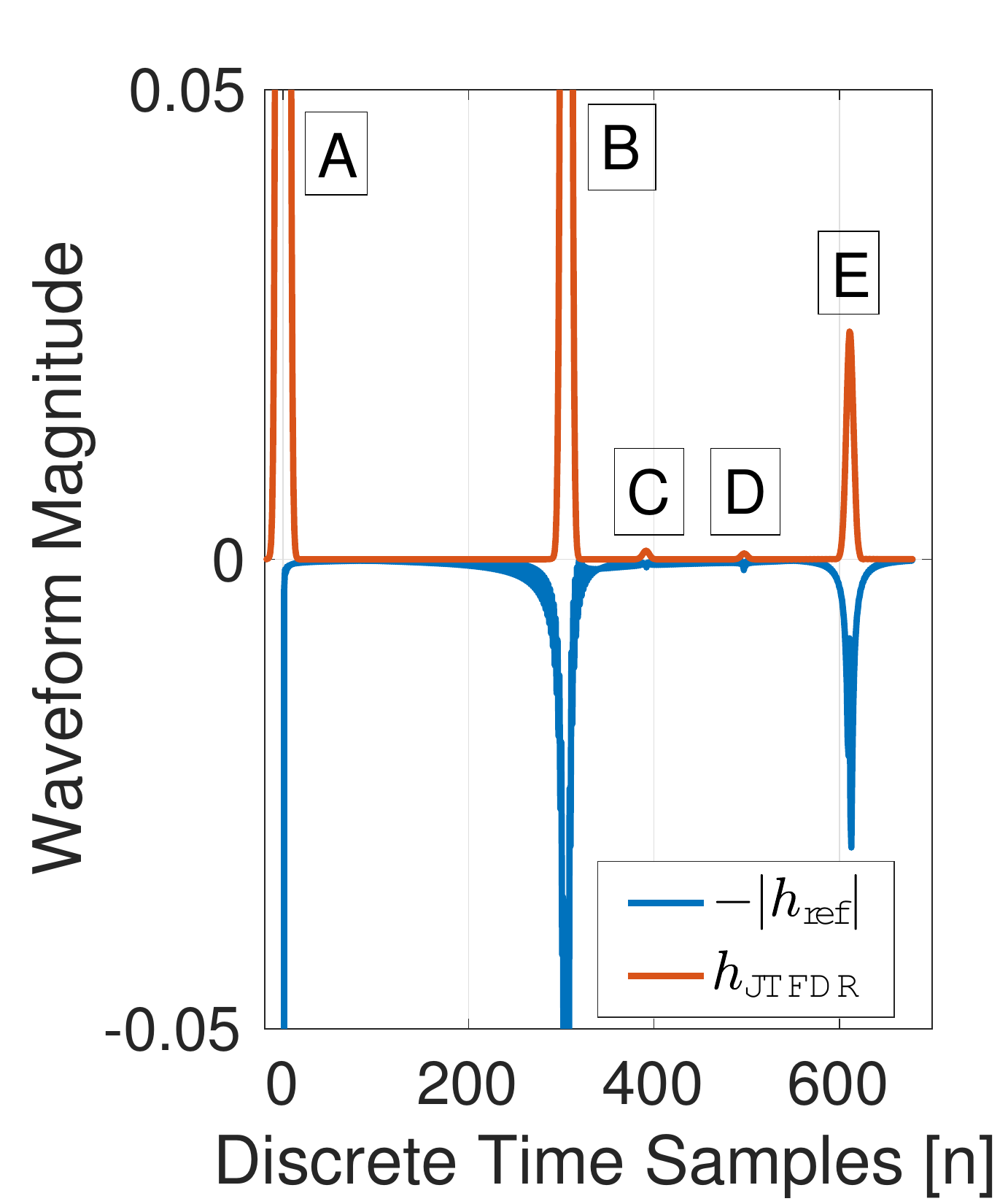} \label{fig:RX_Result_hsi_a}}
     	\subfloat[]{\includegraphics[width=4cm]{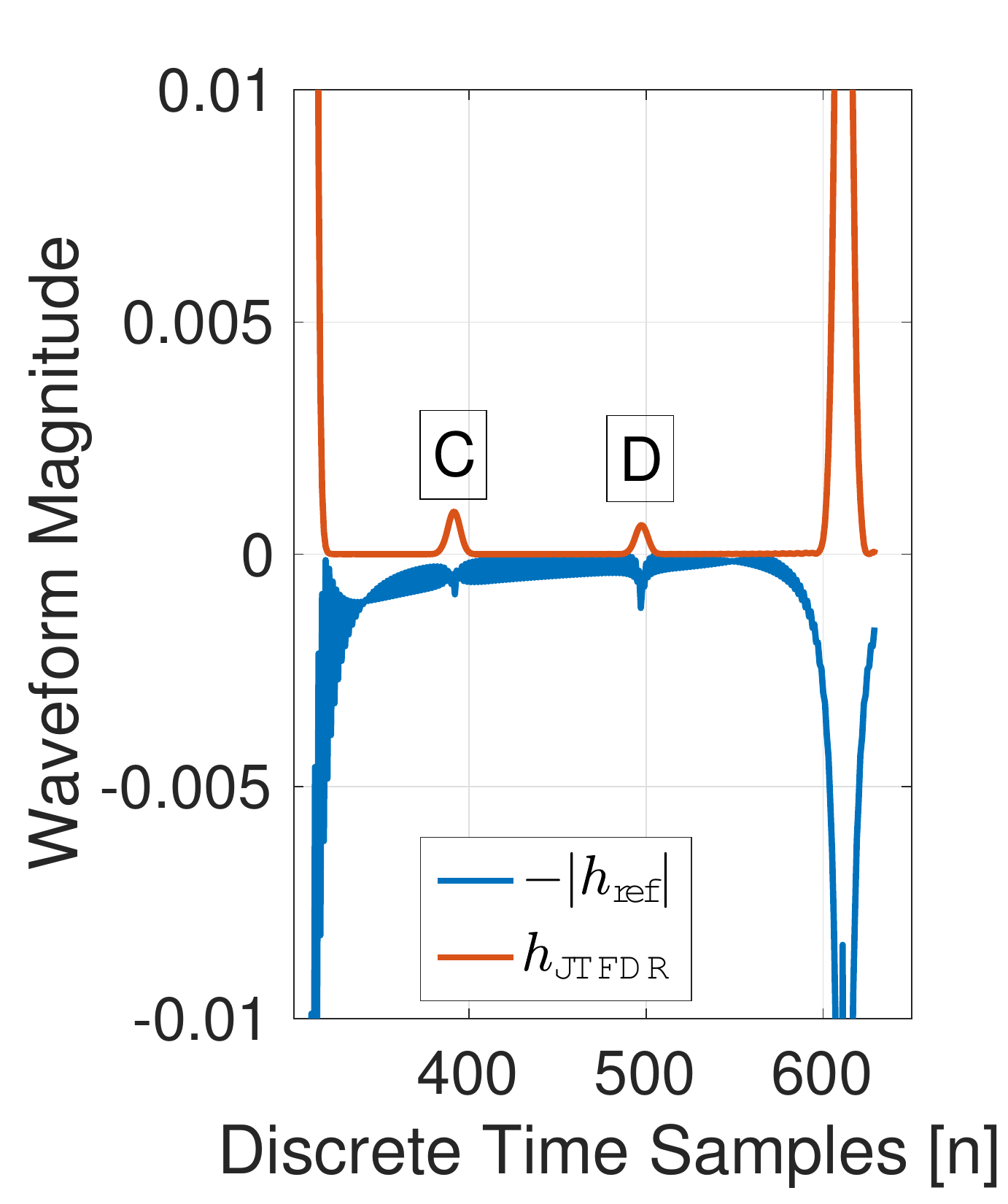} \label{fig:RX_Result_hsi_b}}
	  \vspace*{-5pt}
    \caption{(a) $h_\mathsf{JTFDR}$ and $-|h_\mathsf{ref}|$ {(normalized to the magnitude of peak A)} when a degradation is present between $\RX$ and $\BP$. (b) Zoomed at the degradation locations $C$ and $D$.}
	\vspace*{-.1in}
	 \label{fig:RX_Result_hsi}
\end{figure}

{The peak locations} in Fig.~\ref{fig:TX_Result_hsi} are seen to be at $n_\textsf{A}=0$, $n_\textsf{B}=129$, $n_\textsf{C}=232$, $n_\textsf{D}=305$. Since the network topology is known \textit{a-priori}, i.e., that $\mathsf{BP}$ is $\ell_0=500~\mathrm{m}$ away from $\mathsf{PLM1}$ and {that} $n_\textsf{D}$ indicates the reflection from $\mathsf{BP}$, we can compute the two ends of the LD to be $\ell_0n_\textsf{B}/n_\textsf{D}=211~\mathrm{m}$ and $\ell_0n_\textsf{C}/n_\textsf{D}=380~\mathrm{m}$ apart from $\mathsf{PLM1}$, respectively. While we predict the start location of the degradation accurately, the slight disagreement in locating the degradation end point can be attributed to the slower wave propagation speed in the LD region. For a section of cable with $\epsilon_\mathsf{total}$ (see~\eqref{eqn:EpsilonTotal}), the wave propagation velocity can be computed as 
\begin{equation}
\label{eqn:propagateSpeedExpression}
v=\frac{1}{\sqrt{\mu \epsilon_0 \Re(\epsilon_\mathsf{total})}}. 
\end{equation}
When {$\gamma_\mathsf{local}\gg\gamma_\mathsf{homo}$}, $v$ at the region with a localized WT degradation is noticeably different from other regions. This is also shown in~\cite{mugala2005high, livani2014machine, forstel2017grid}, where fault or degradation localization in severely aged service cables yields significant deviance from their true values. 

\begin{figure}
     \centering
     	\subfloat[][]{\includegraphics[width=4cm]{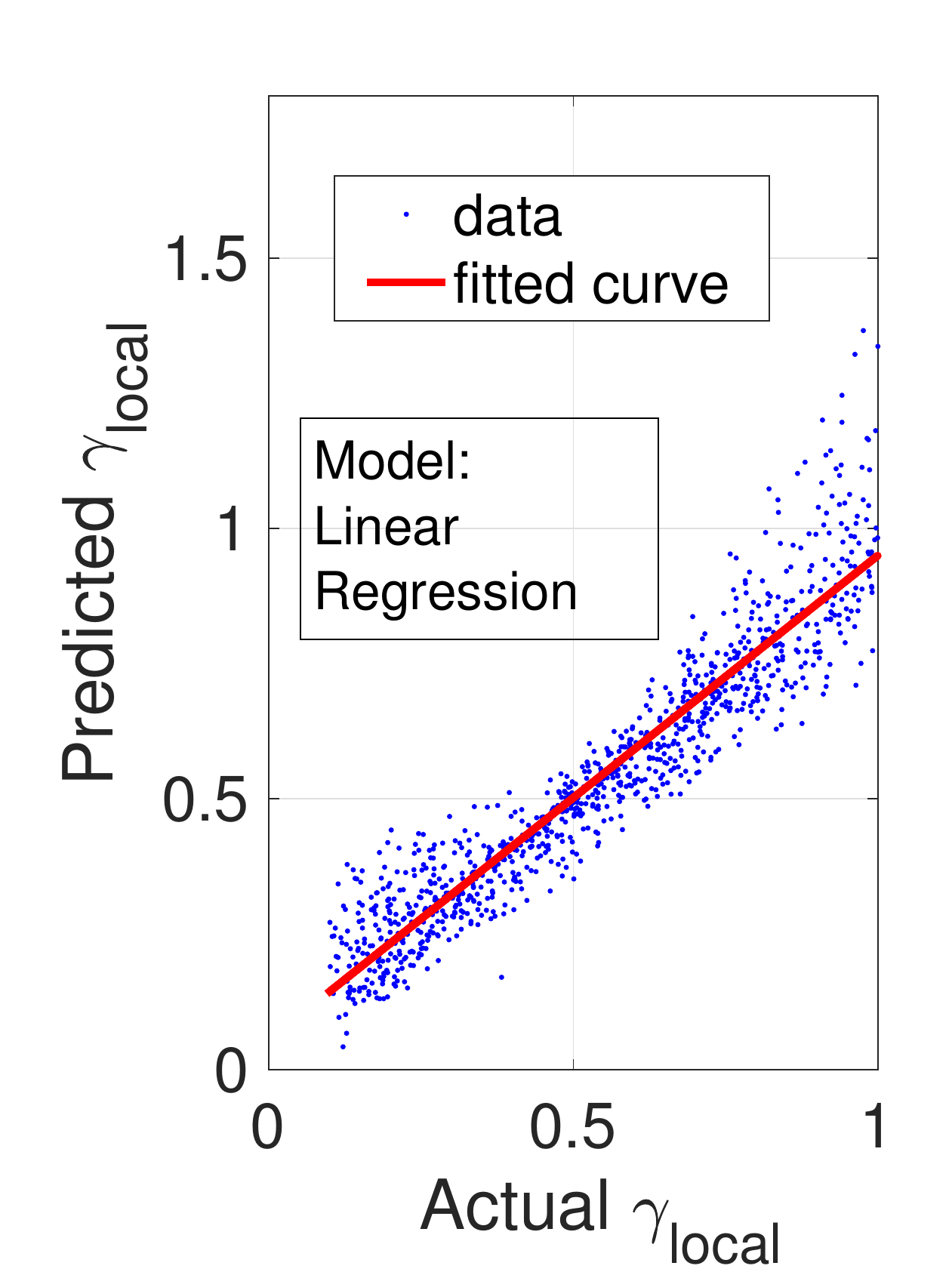} \label{fig:LR_Result1}}
     	\subfloat[][]{\includegraphics[width=4cm]{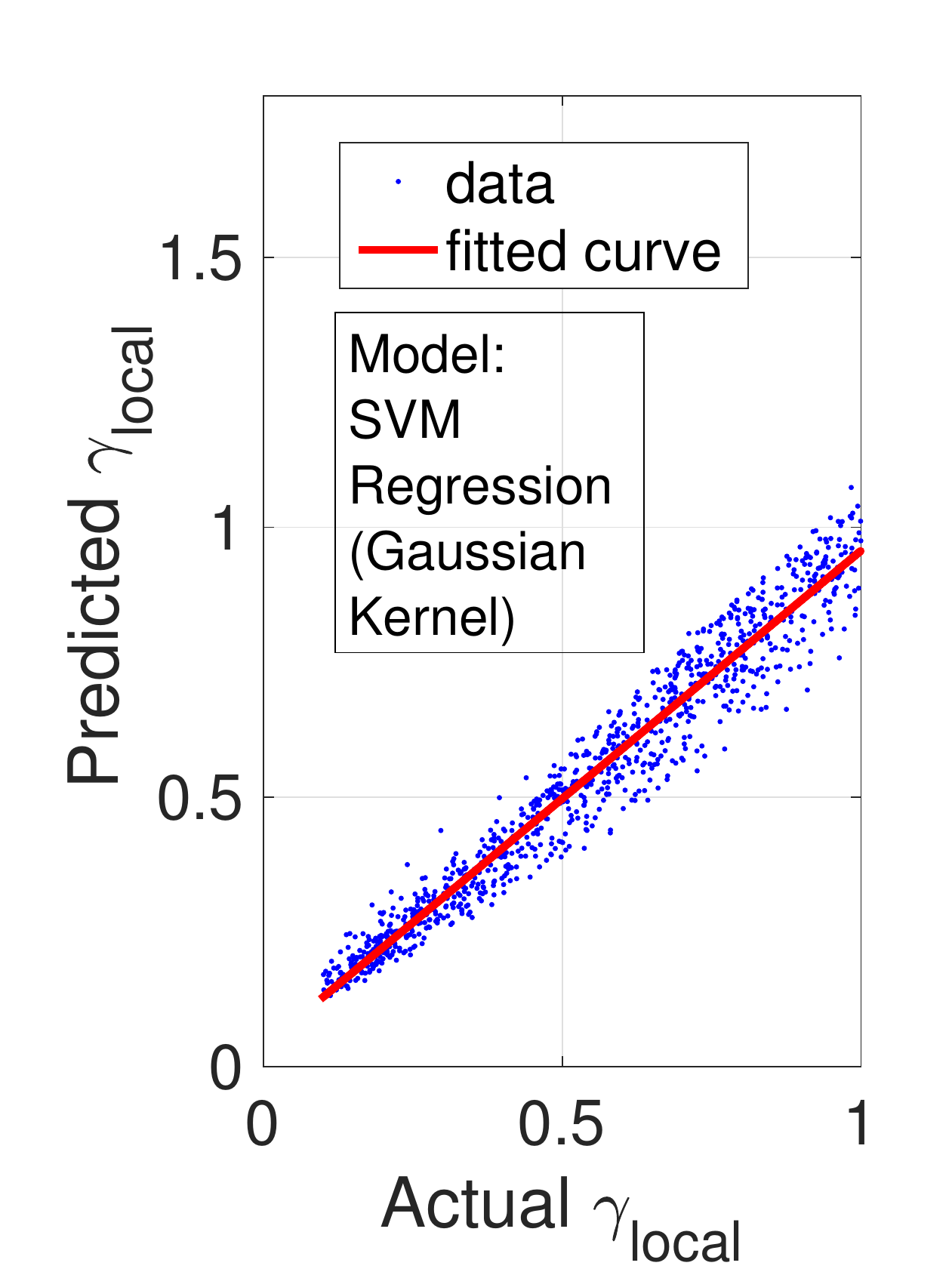} \label{fig:LR_Result2}}
     	\caption{Prediction of $\gamma_\mathsf{local}$ using peak magnitudes and locations of the PLM-JTFDR waveforms. (a): linear regression. (b): SVM regression with Gaussian kernel.}
	\vspace*{-10pt}
	 \label{fig:LR_Result}
\end{figure}

Our evaluations thus far show the capability of our proposed PLM-JTFDR in identifying and locating an LD. However, this only completes a portion of the LD diagnostics tasks. In the following, we attempt to perform the remaining LD severity estimation by exploiting the principle that a more severe degradation causes stronger discontinuities in cable insulation dielectric properties, and thus results in a higher reflection peak in $h_\mathsf{JTFDR}$. Therefore, we monitor the peak amplitudes to infer the extent of degradation. However, we find the relationship between peak amplitudes and the degradation severities to be non-linear, and furthermore affected by signal attenuations along the line and changing load conditions. This can be observed from Fig.~\ref{fig:LR_Result}, where we apply regression techniques in an attempt to predict the degradation severity level of an LD. We train our machine with $3600$ aging profiles using the peak amplitudes and locations, and their associated $\gamma_\mathsf{local}\sim\mathscr{U}(0.1,1)$. Our prediction results $1000$ other agind profiles subject to LD with $\gamma_\mathsf{local}\sim\mathscr{U}(0.1,1)$ is shown in Fig.~\ref{fig:LR_Result}.


We observe in Fig.~\ref{fig:LR_Result}(a) that applying linear regression to predict $\gamma_\mathsf{local}$ provides an unsatisfactory prediction performance. On the other hand, the results in Fig.~\ref{fig:LR_Result}(b) are more encouraging, where the performance is improved by using a support vector machine (SVM) regression using the Gaussian kernel. This further motivates us to include the results of PLM-JTFDR in some form within an ML-based cable diagnostics solution.


\subsection{{A Machine Learning Framework}}
\label{subsec:machine_learning}

\begin{figure}[!t]
	\centering
	\includegraphics[width=8cm]{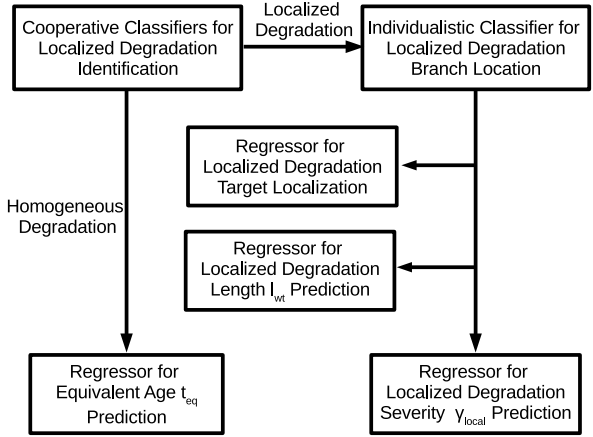}
	\caption{Proposed ML framework for cable health monitoring.}
	\label{fig:ageing_inference_ML}
\end{figure}

Our proposed ML-based cable diagnostics solution is shown in Fig.~\ref{fig:ageing_inference_ML}. We formulate the problems of aging profile type identification and branch localization as supervised classification, and the problems of degradation localization and aging severity assessment as regression. {For each of the supervised classification tasks, we use the presence or absence of an LD as the training label. For every supervised regression task, the value to be predicted, e.g., $\gamma_\mathsf{homo}$ and $\ell_\mathsf{wt}$, as shown in Fig.~\ref{fig:ageing_inference_ML}, is used as the training label.}

Throughout our analysis, we consider the generic $T$-network topology of Fig.~\ref{fig:networkTopology}. To determine the presence of an LD, we run classifiers in each of the PLMs to identify an LD in their nearest branch(es). We then employ a cooperative technique, by which we conclude that no LD exists if all of the PLMs report the absence of an LD at their nearest branches. In such a case, the cables are predominantly subject only to homogeneous degradation across the network. On the other hand, when a PLM reports the presence of an LD, we face the challenge of topological ambiguity, where the degradation could be located on a branch in either direction of the PLM\footnote{For our evaluations, we address the condition where only one LD is present in the considered network. We believe that this assumption is not far from reality since we expect multiple LDs to occur sequentially, and therefore can be addressed one at a time.}. In the network of Fig.~\ref{fig:networkTopology}, when the $i$th PLM, $p_i$, reports an LD, the LD can lie either between  $p_i-\BP$ or $p_i-\BE_i$. To eliminate this ambiguity, we perform a second round of classification to detect the LD branch location. In the second round, we let $p_j$ ($j \in \{1,2,3\},  j \neq i$) confirm whether the degradation resides between $p_i$ and $\BP$ using a similar classification procedure. Once the presence of an LD is confirmed within the considered network, we use different ML regressors in the PLM closest to the LD for severity assessment and localization. In particular, we use an ML regressor to find the associated $\gamma_\mathsf{local}$, another one to predict its length, i.e., $\ell_\mathsf{wt}$, and a final one to predict the distance between the PLM and the near-end of the LD, which we also refer to as the \textit{target} location.

On the other hand, when the aging profile is identified to be homogeneous, we use an ML regressor to predict an equivalent cable age, $t_\mathsf{eq}$, which provides an intuitive indication into the overall degradation severity of the evaluated cable insulation. Specifically, we train a machine to predict $\gamma_\mathsf{homo}$ of the evaluated homogeneous aging profile and use~\eqref{eqn:CableAge} to compute $t_\mathsf{eq}$ with the nominal cable parameters listed in Table~\ref{table:parameters}.


\subsection{Machine Learning Algorithms}
\label{subsec:machine_learning}

For both classification and regression, we use two sets of ML techniques, namely SVM and boosting, following their success both in previous cable diagnostics evaluations as well as in other domains~\cite{huo2018grid, forstel2017grid, murphy2012machine}. SVM is a classical and popular ML technique, which constructs support vectors of hyperplanes, from a subset of the training data, for predictions. It encodes sparsity in the hinge loss function, which is a convex upper-bound for the non-smooth $0-1$ loss function, and results in a small number of support vectors compared to the number of training samples~\cite[Ch. 14]{murphy2012machine}. The sparsity and the well-known large margin principle allows SVM to provide accurate predictions for unseen data samples. Furthermore, optimization problems in SVM operation can be efficiently solved by convex quadratic programming, and a high-dimensional space can be explored by kernel tricks.

Boosting is one of the meta-ML algorithms that works by consolidating multiple weak learners into a strong learner. It applies the weak learners sequentially to weighted versions of the data, where a higher weight is allocated to examples that suffered greater inaccuracy in earlier prediction rounds~\cite[Ch. 16]{murphy2012machine}. These weak learners are typically only marginally better than random guessing but are computationally simple. Boosting is also known to be robust to over-fitting, and can be efficiently executed since it is a forward stage-wise additive model~\cite[Ch. 16]{murphy2012machine}. We use the adaptive boosting (AdaBoost) for classification and the gradient boosting with square loss function, known as least-square boosting (L2Boost), for regression.

Apart from the specific choices of the algorithms, the performance of our solution also depends on the number of samples used to train the machines, $n_\mathsf{TR}$. We detail our choice of $n_\mathsf{TR}$ as well as the number of testing samples, $n_\mathsf{TE}$, in Appendix~\ref{sec:sample_number}, followed by a discussion of our strategies to prevent over-fitting of our model in Appendix~\ref{sec:avoid_overfitting}.

\subsection{Feature Selection}
\label{sec:feature_selection}

For each of the ML tasks, selecting the feature set is crucial for a successful ML performance. We have shown in our prior works~\cite{huo2018grid, forstel2017grid} that cable degradations cause higher dielectric losses and thus greater attenuation in $H_\f$ across the frequency band. Therefore, we include the $m$th-order moments ($m \in \{1,2,3,4\}$) of $|H_\f|$ in our feature library to train our machine. Additionally, various studies in the literature, e.g.,~\cite{mugala2005high, livani2014machine, forstel2017grid}, have shown that degradations reduce the velocity of wave propagation in the degraded section of the cable. Thus, we also include the locations and amplitudes of the peaks in the channel impulse response, $h_\f$, and the $m$th-order moments ($m \in \{1,2,3,4\}$) of $\angle{H_\f}$ in our feature library. Similarly, we also include the $m$th-order moments ($m \in \{1,2,3,4\}$) of $|H_\mathsf{ref}|$ and $\angle{H_\mathsf{ref}}$ in our feature library to provide additional insights for cable health monitoring.

Further, our evaluation results in Figs.~\ref{fig:TX_Result_hsi}-\ref{fig:LR_Result} have also shown that PLM-JTFDR provides cleaner waveforms for extracting features for {degradation identification and localization,} and also indicative information in predicting the LD severity. Therefore, we also extract peak locations and magnitudes from the resultant PLM-JTFDR waveforms as features for our ML operation.


\section{Simulation Results}
\label{sec:numerical}
In this section, we demonstrate the effectiveness of our ML solution through numerical evaluations for {automated cable health condition monitoring.} Throughout our simulations, we apply the network topology shown in Fig.~\ref{fig:networkTopology}, and the cable aging model and PLC channel characterization described in Section~\ref{sec:ageingModel} and Section~\ref{sec:plcModel}, respectively. In addition, we set $\gamma_\mathsf{homo}\sim\mathscr{U}(0,0.05)$, $\gamma_\mathsf{local}\sim\mathscr{U}(0.1,1)$, $\ell_\mathsf{WT}\sim\mathscr{U}(100,300)~\mathrm{m}$, and the center of the localized WT degradation randomly located within $100~\mathrm{m}$ from the center of a branch in Fig.~\ref{fig:networkTopology}. Since $n_\mathsf{TR}$ and $n_\mathsf{TE}$ vary for each of the ML tasks, we specify our chosen tuple $(n_\mathsf{TR}, n_\mathsf{TE})$ with all our results.

\subsection{Cooperative LD Identification}
\label{subsec:branch_location_s1}

\begin{figure}[!t]
	\centering
	\includegraphics[width=7cm]{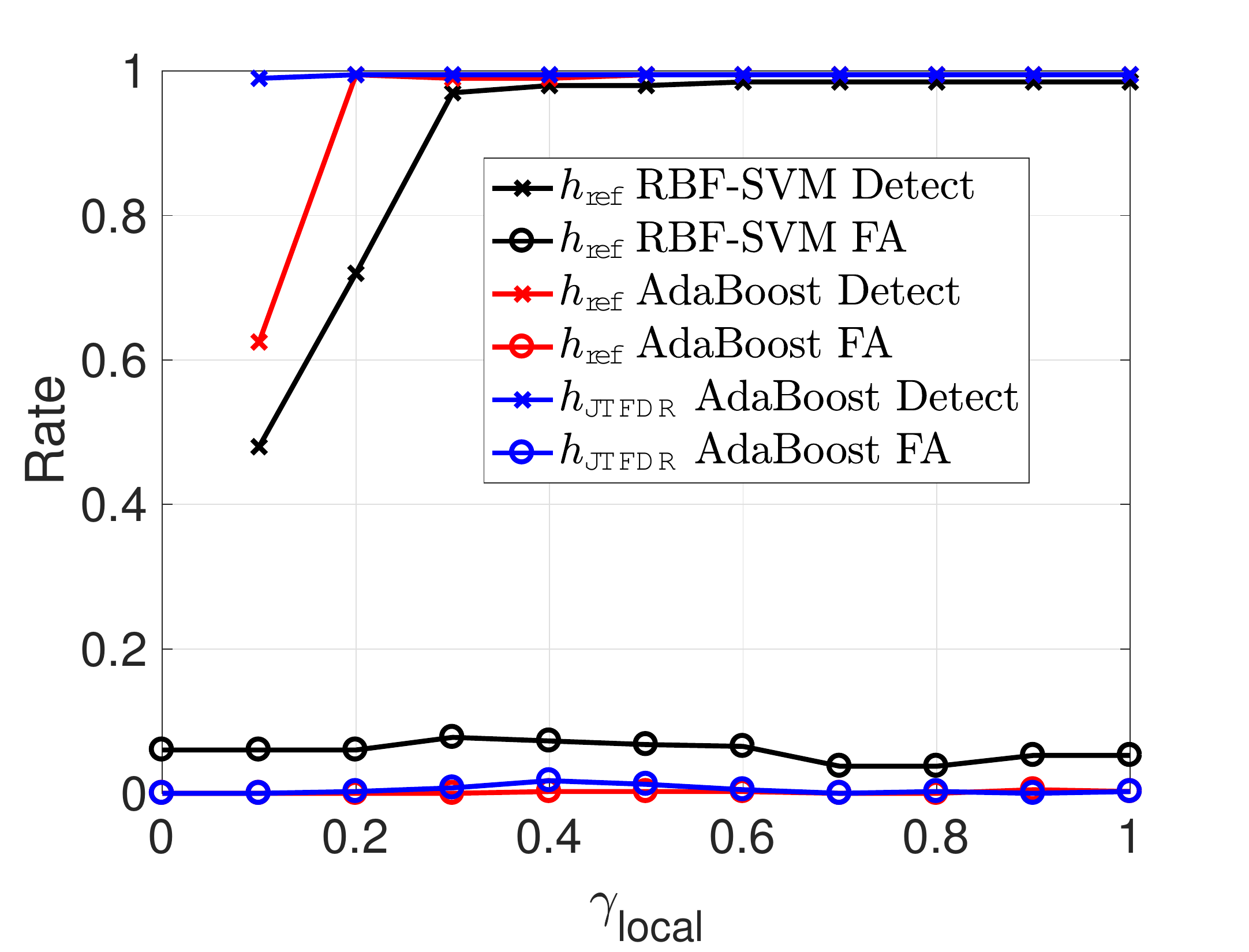}
	\caption{Detection and FA rates for Cooperative LD Identification, with $(n_\mathsf{TR}, n_\mathsf{TE})=(7000,6600)$.}
	\label{fig:branch_location}
\end{figure}

We use the $i$th PLM, $p_i$, ($i \in \{1,2,3\}$) to identify the presence of a localized WT degradation, regardless of whether it resides between $p_i-\BP$ or $p_i-\BE_i$. The result of this exercise {for varying} $\gamma_\mathsf{local}$ is shown in Fig.~\ref{fig:branch_location}. 
We observe in  Fig.~\ref{fig:branch_location} that using $h_\mathsf{JTFDR}$ to extract features provides near-perfect detection with negligible FA rates across all values of $\gamma_\mathsf{local}$, while using $h_\mathsf{ref}$ provides unsatisfactory detection especially at lower $\gamma_\mathsf{local}$, with either SVM with radial basis function (RBF) kernel or the adaptive boosting (AdaBoost) {as our considered} ML algorithms.
The improved performance is expected due to the nature of $h_\mathsf{JTFDR}$ in comparison to that of $h_\mathsf{ref}$, i.e., more prominent peaks and a smoother floor in $h_\mathsf{JTFDR}$, as discussed in Section~\ref{subsec:PLM_JTFDR}. Therefore, the result of Fig.~\ref{fig:branch_location} shows that the detection performance can be significantly improved by adopting PLM-JTFDR into the ML framework. 

\subsection{LD Branch Location}
\label{subsec:branch_location_s2}

\begin{figure}
     \centering
     	\subfloat[]{\includegraphics[width=4cm]{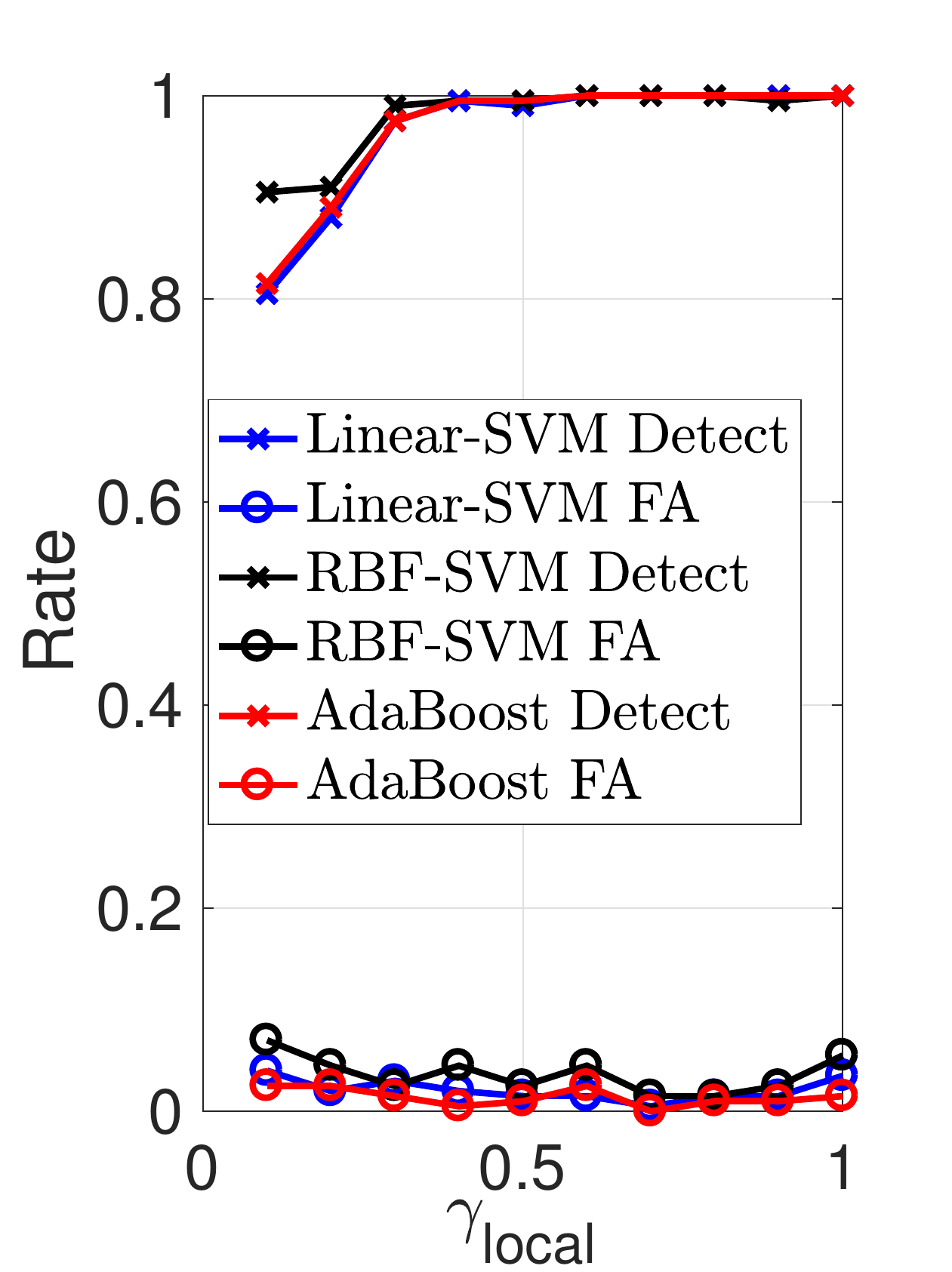} \label{fig:branch_location_s2_a}}
     	\subfloat[]{\includegraphics[width=4cm]{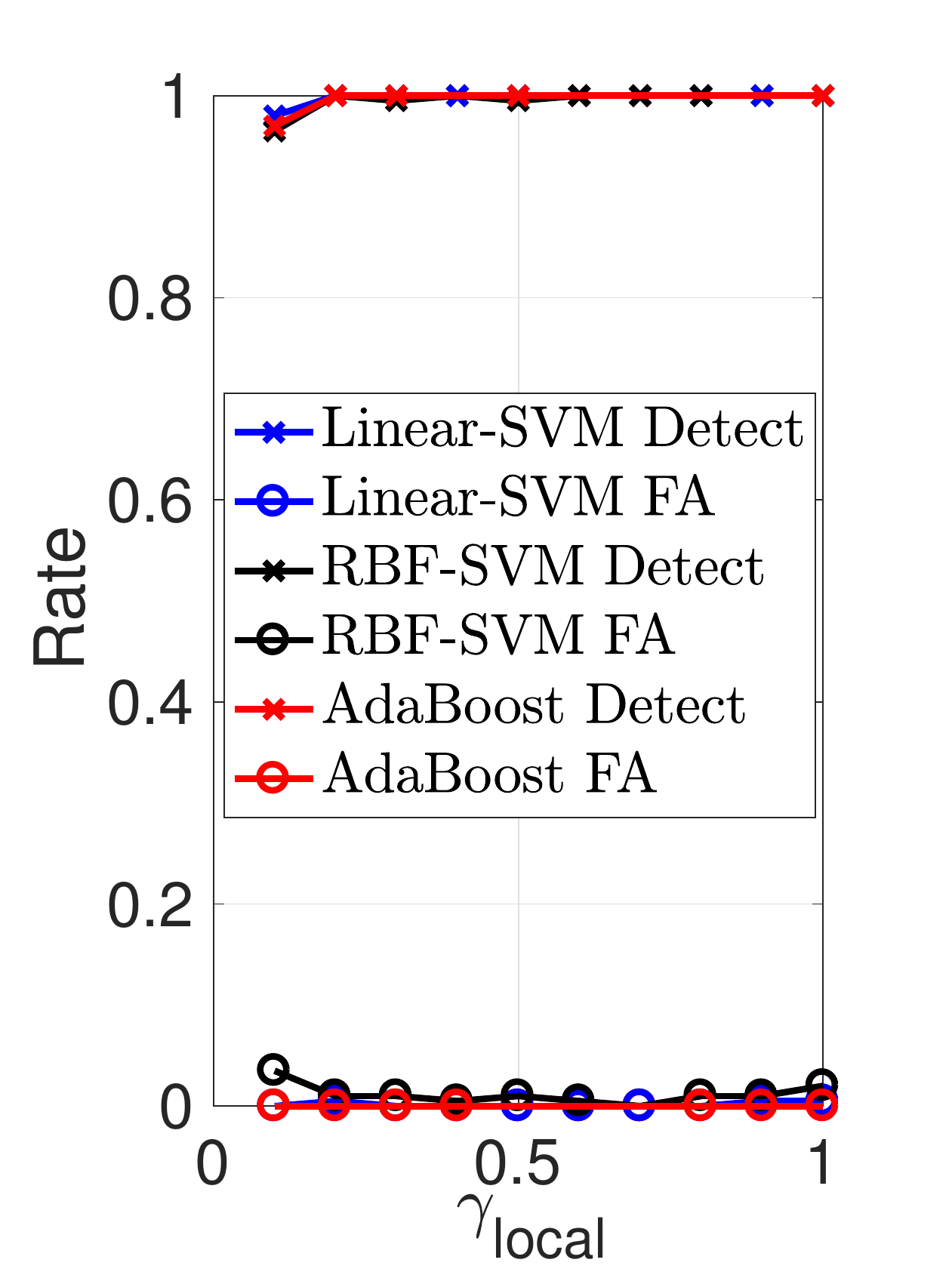} \label{fig:branch_location_s2_b}}
	 \vspace*{-5pt}
     	\caption{Detection and FA rates for LD Branch Location with (a) the feature set used for Fig.~\ref{fig:branch_location}, and (b) with additional features for improved results, both of which are obtained with $(n_\mathsf{TR}, n_\mathsf{TE})=(4000,4000)$.}
	\vspace*{-10pt}
	 \label{fig:branch_location_s2}
\end{figure}

Once a localized WT degradation is detected between $p_i-\BP$ or $p_i-\BE_i$, we let $p_j$ ($j \in \{1,2,3\},  j \neq i$) confirm whether the degradation resides between $p_i$ and $\BP$. The detection results for this phase is shown in Fig.~\ref{fig:branch_location_s2}.
At first, we use the same feature set for this task as that used in the cooperative LD identification. We observe in Fig.~\ref{fig:branch_location_s2}(a) that this results in detection rates that are less than desirable, especially for lower values of $\gamma_{\mathsf{local}}$. However, by adding additional features, such as the variance of $h_{\mathsf{JTDFR}}$, the detection rates can be substantially improved across all values of $\gamma_{\mathsf{local}}$, as seen in Fig.~\ref{fig:branch_location_s2}(b).

\subsection{Homogeneous Aging Severity Assessment}
\label{subsec:age_prediction}

\begin{figure}
     \centering
     	\subfloat[]{\includegraphics[width=4cm]{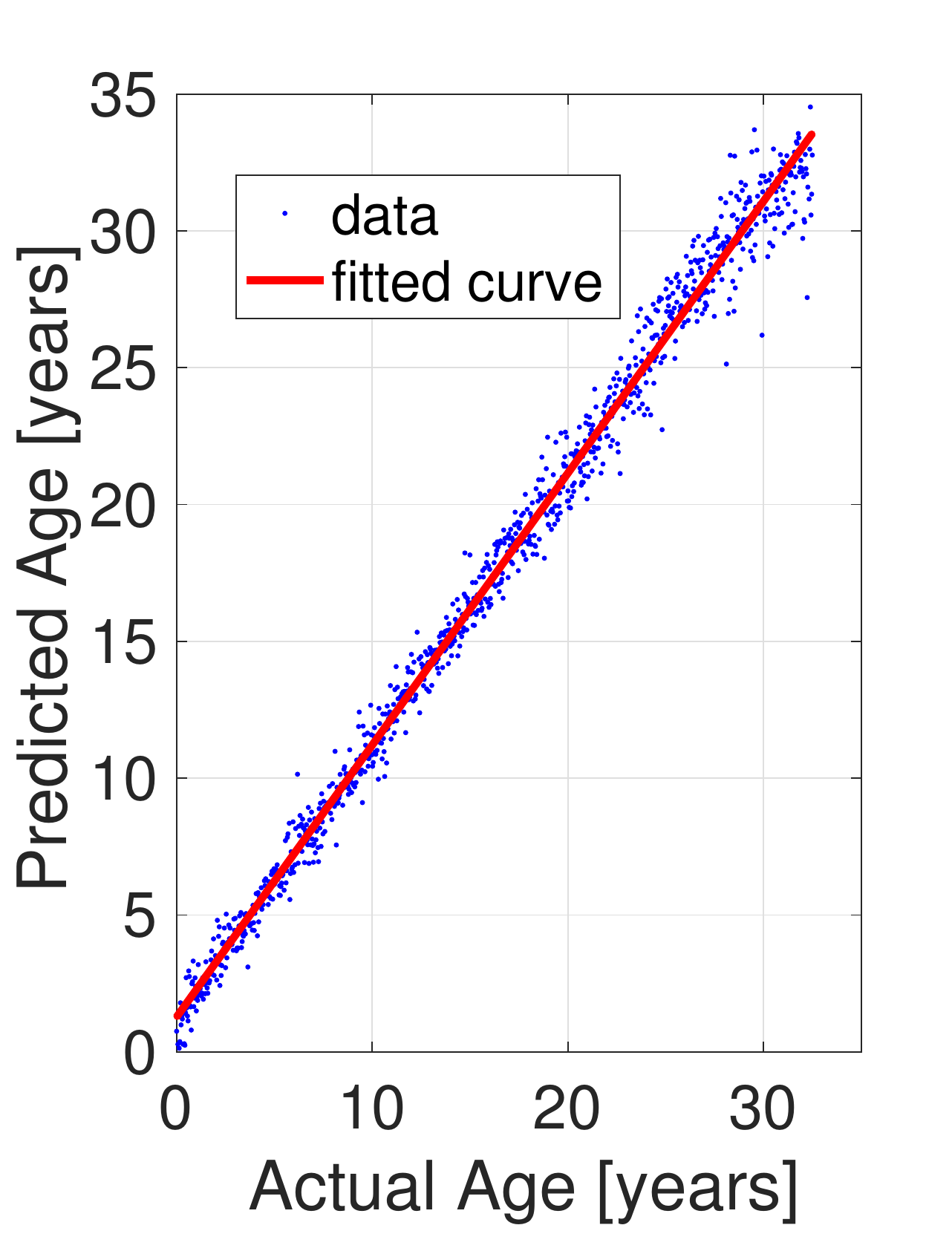} \label{fig:equivalent_age}}
     	\subfloat[]{\includegraphics[width=4cm]{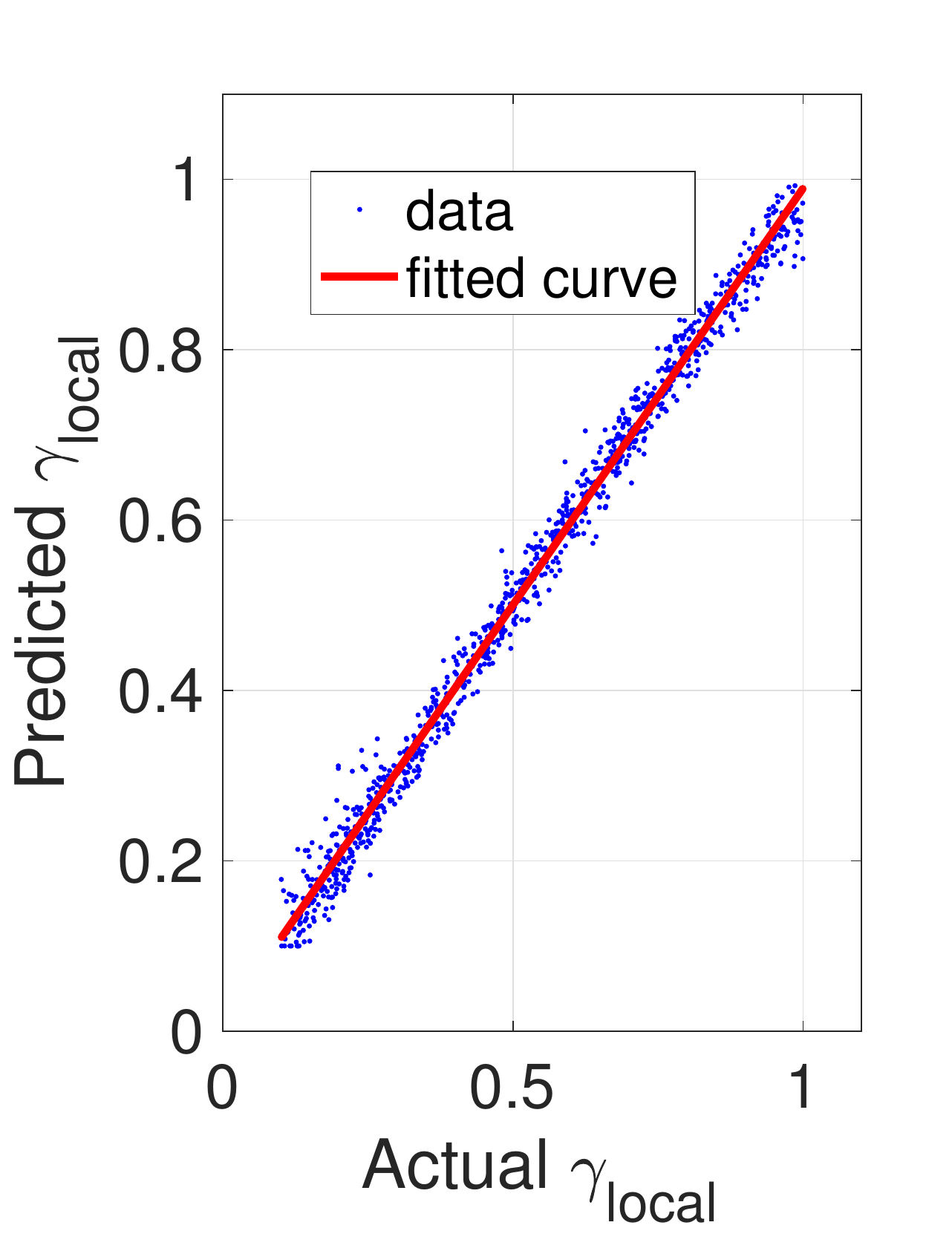} \label{fig:predict_gamma_local}}
	 \vspace*{-5pt}
     	\caption{Degradation severity assessment with $(n_\mathsf{TR}, n_\mathsf{TE})=(3600, 1000)$ for (a) homogeneous aging and (b) LD.}
	 \label{fig:predict_severity}
\end{figure}

When the aging profile type classification determines {a homogeneous aging profile}, we use L2Boost to train a machine for the prediction of the degradation severity level. We {train and test} with different homogeneous degradation severities quantified by their equivalent age as computed in~\eqref{eqn:CableAge}. To emulate realistic degradations, we limit $t \sim \mathscr{U}(0,32.5)~\mathrm{years}$ in our simulations{~\cite[Ch. 6]{gill2008electrical}}. {The equivalent age prediction performance is shown in Fig.~\ref{fig:predict_severity}(a).} We observe that the estimated age closely matches the actual age. Furthermore, the prediction accuracy is considerably improved when compared to the state-of-the-art~\cite{huo2018grid}, as we now use not only $H_\f$, {but also $H_\mathsf{ref}$}.

\subsection{LD Severity Assessment}
\label{subsec:predict_gamma_local}

Next, we consider the condition where the classifier indicates the presence of an LD, to assess its severity. {To this end, we train an LSBoost regressor with a degradation randomly located anywhere in the $\TX-\BP$ branch, since we rely on the node closest to the degradation for assessment. The prediction results are presented in Fig.~\ref{fig:predict_severity}(b).} Similar to the performance gains seen in Section~\ref{subsec:age_prediction}, our results show significant improvement in prediction accuracy when compared to the state-of-the-art~\cite{huo2018grid}, due to the additional insight obtained from the waveform of PLM-JTFDR and other features extracted from $H_\mathsf{ref}$.

\subsection{LD Location}
\label{subsec:location}
\begin{figure}
     \centering
     	\subfloat[]{\includegraphics[width=4cm]{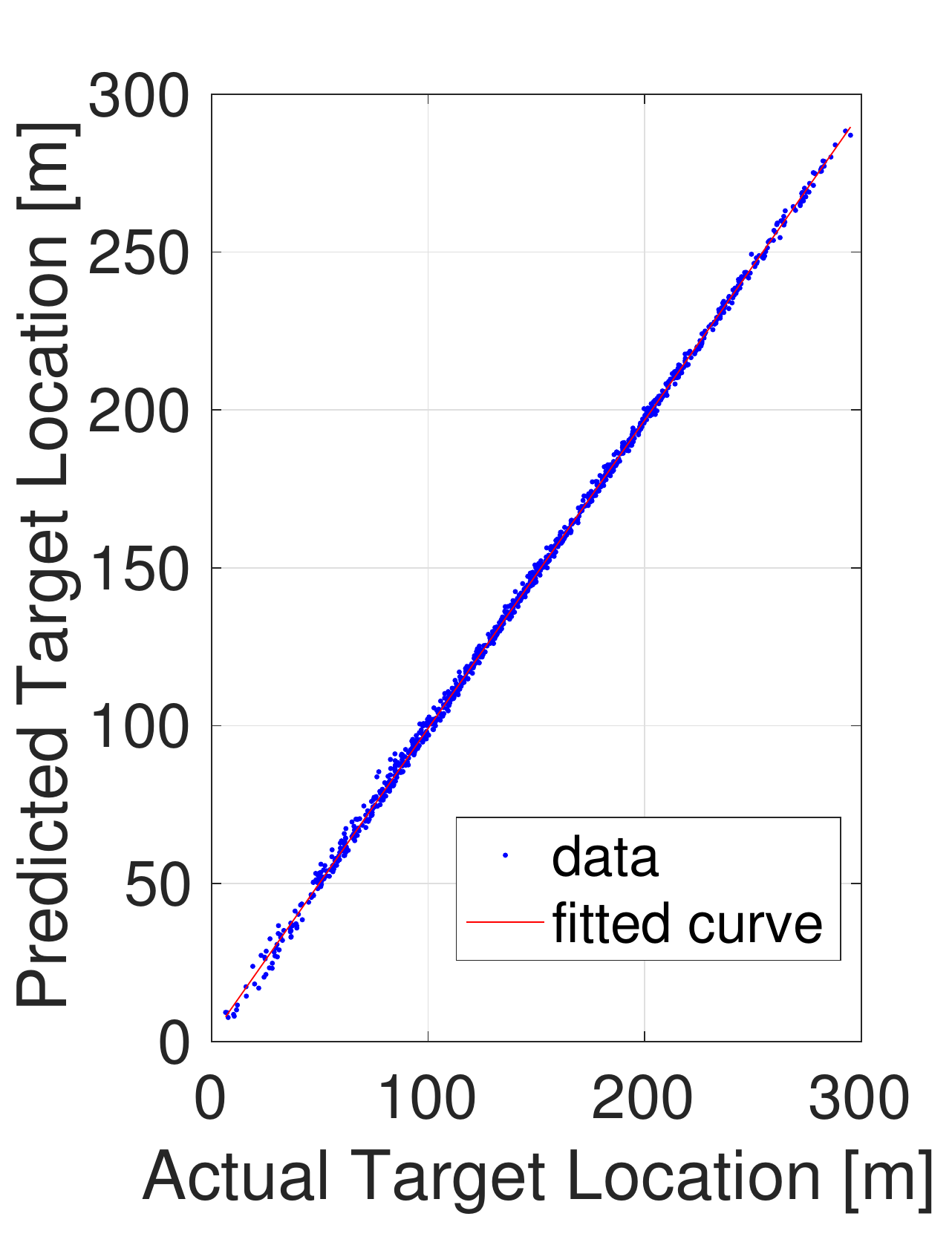} \label{fig:target_location}}
     	\subfloat[]{\includegraphics[width=4cm]{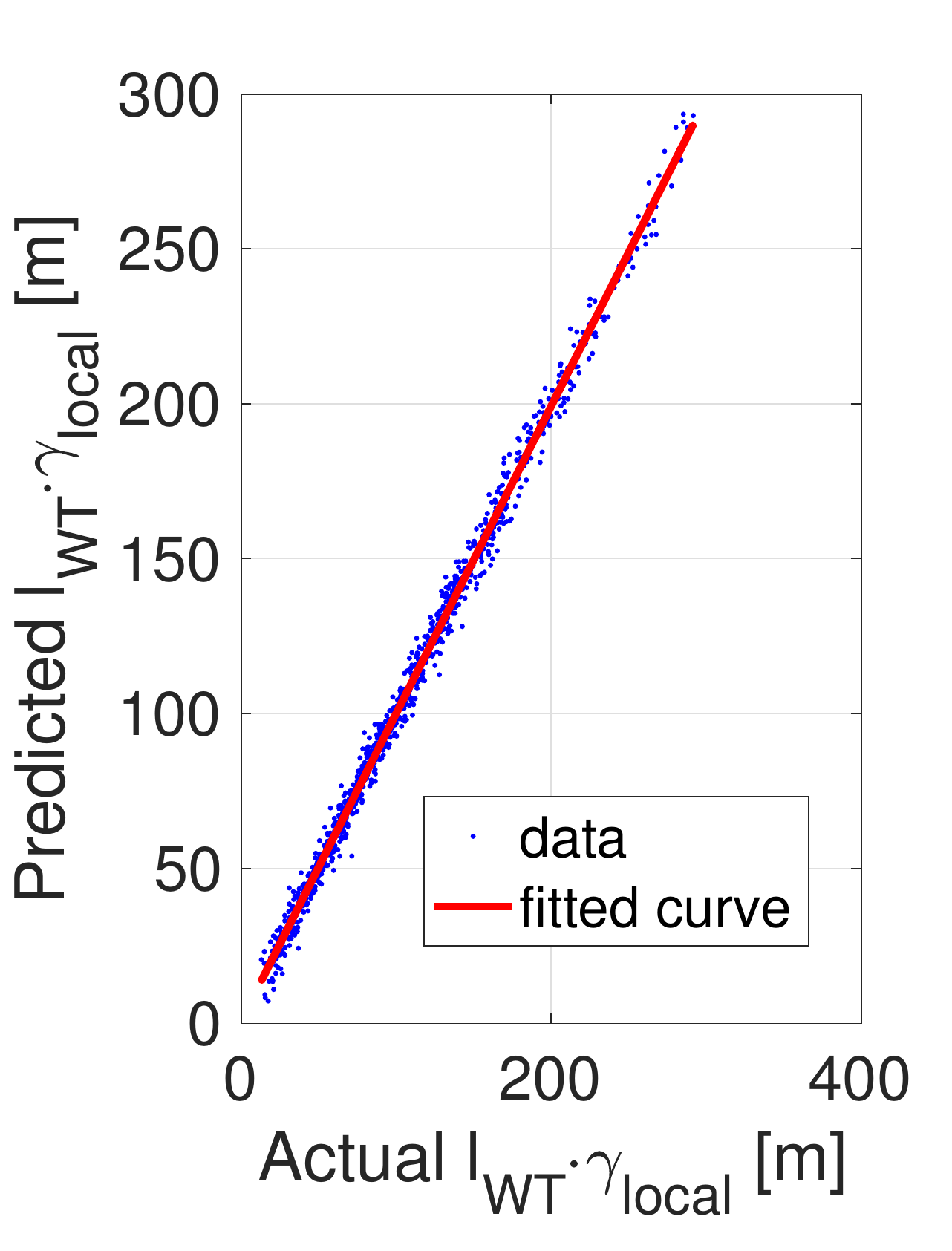} \label{fig:predict_length}}
     	\caption{LD location results with $(n_\mathsf{TR}, n_\mathsf{TE})=(3600, 1000)$ for predicting (a) the target location and (b) the degradation length.}
	 \label{fig:localization_results}
\end{figure}

Thus far, we have predicted the type of degradation present and its severity. As our final diagnostics stage, we attempt to locate the salient localized cable degradation so that further efforts in preventing an in-service fault can be concentrated. Locating an LD consists of determining the positions of its two ends. For this purpose, we follow the procedure of first determining its starting point, or the \textit{target} point, and then estimating the degradation length. {For the localization of the target point, we train an SVM for regression with linear kernel.} The results of predicting the target location can be seen in Fig.~\ref{fig:localization_results}(a). We notice that the prediction accuracy is near perfect with negligible variance of individual predictions from the fitted curve. This accurate target location prediction is made possible due to the peak location clarity in $h_{\mathsf{JTFDR}}$.

Next, we predict the length of the degradation to identify its end point. However, we found that the length prediction was highly inaccurate with any ML algorithm or any combinations of extracted features.
Therefore, we device a workaround to this challenge by instead predicting the product $\ell_\textsf{WT} \cdot \gamma_\textsf{local}$, which can be predicted with high accuracy, and then using our previously predicted values of $\gamma_\textsf{local}$ (in Fig.~\ref{fig:predict_severity}(b)) to determine $\ell_\textsf{WT}$. The prediction results for $\ell_\textsf{WT} \cdot \gamma_\textsf{local}$ is shown in Fig.~\ref{fig:localization_results}(b), where we clearly notice that the fitted curve of the predictions is a straight line with unit slope and nearly passing through the origin, thereby confirming the high accuracy of our results. 

\section{Discussion} \label{sec:discussion}
In this section, we provide a brief discussion on our proposed solution by comparing it with prior arts and exploring its robustness in non-ideal settings.

\subsection{Related Work}
\subsubsection{Reflectometry Methods}

{Utilities typically use reflectometry methods, such as TDR or JTFDR, for detecting faults and degradations on the line~\cite{shi2010detection, burkes2015water, lee2013non, wang2010application}. While our solution includes the use of the same underlying principle to enhance our proposed ML framework, we synthesize the results of JTFDR using the already existing PLMs that inherently estimate the reflection channel. Therefore, no additional dedicated components are required as in conventional reflectometry methods. Further, by incorporating the reflectometry methods into a ML framework, we are able to comprehensively {infer} the cable health {conditions}, which conventional methods are unable to achieve.}

\subsubsection{Fault localization}\label{subsec:fault}
Several prior works have proposed and developed solutions to use PLMs for fault localization, for e.g.,~\cite{lazaropoulos2017main, lehmann2016diagnostic, passerini2017power, passerini2017full}. Apart from these works facing practical limitations, such as, say, requiring a reference CFR measurement for any given load condition, detecting the presence of a fault on the main line is a reactive response to a service failure. In contrast, we develop solutions in our work to estimate cable {degradations} in order to take preventive measures to avoid a fault. Furthermore, assessing cable degradations is also non-trivial when compared to fault detection as the signal reflections tend to be more obscure~\cite{griffiths2006invisible}, also as seen in Fig.~\ref{fig:TX_Result_hsi} and Fig.~\ref{fig:RX_Result_hsi}.

\subsubsection{ML Techniques}
ML-based \textit{data-driven} methods have also been previously used for cable diagnostics, albeit without using PLC. \cite{ntalampiras2016fault} provides a substantial literature review on fault diagnosis using data-driven methods. Further, \cite{livani2014machine} and~\cite{jiang2011hybrid} also provide techniques to use SVM for cable diagnostics. However, similar to the methods described in Section~\ref{subsec:fault}, these solutions predominantly focus only on fault {diagnostics}, and also do not involve using PLMs.

\subsubsection{Alternative Components for Diagnostics}
Fault identification and localization methods have been proposed in the literature by using phasor measurement unit (PMU) measurements~\cite{gopakumar2015adaptive, jamei2018low, farajollahi2018locating}. While PMUs are traditionally installed in transmission networks for grid status monitoring, use of distribution-level PMUs or micro-PMUs ($\mu$PMUs) has been proposed for distribution networks as well~\cite{farajollahi2018locating, mohsenian2018distribution}. Nevertheless, the benefits of our solution over such methods are evident. {Apart from the added installation cost, the sampling rates and the number of installed $\mu$PMUs} are insufficient to perform degradation localization and severity assessment with the accuracy that we have demonstrated in Section~\ref{sec:numerical}.



\subsection{Robustness to Non-idealities}

\begin{figure}
     \centering
     	\subfloat[]{\includegraphics[width=4cm]{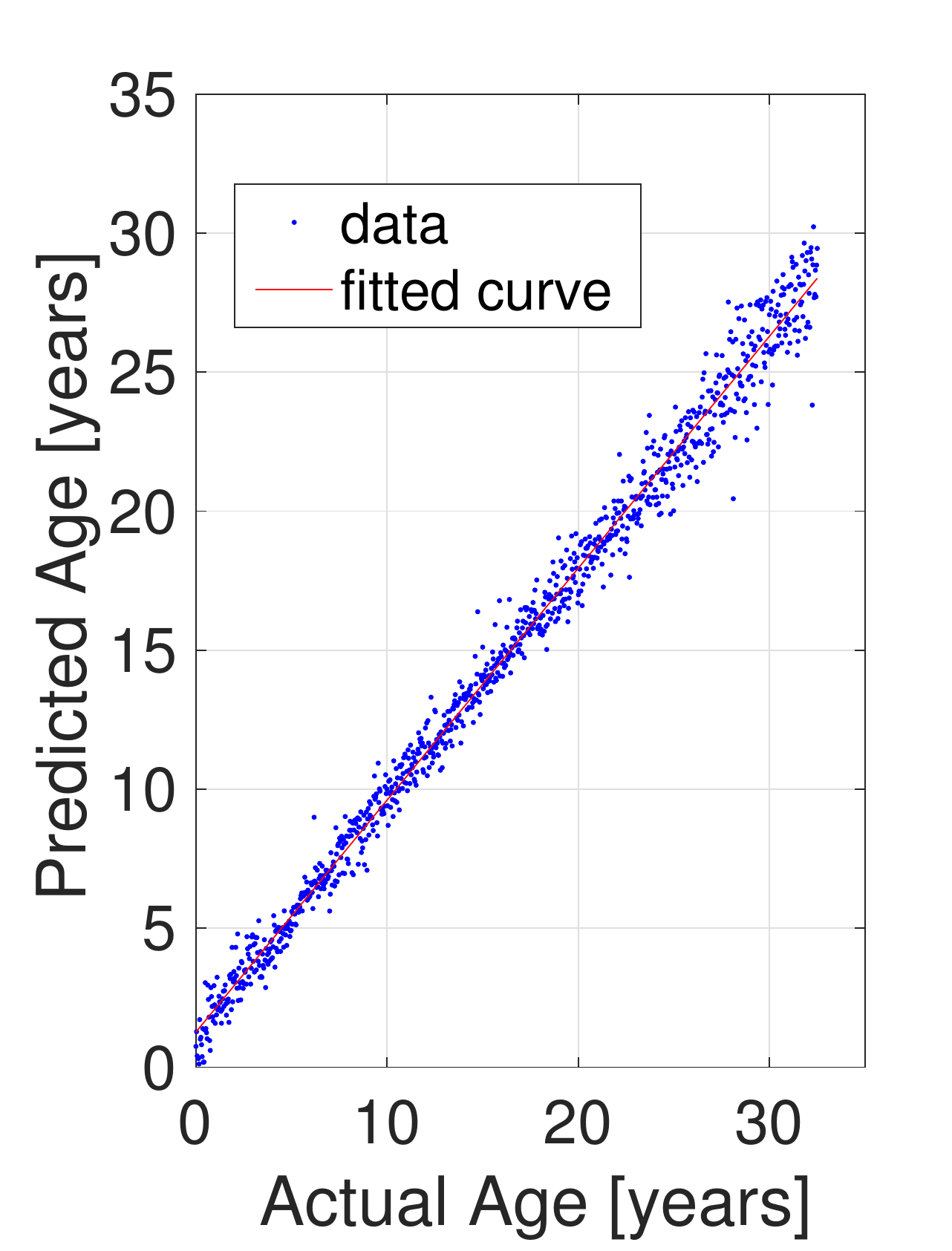} \label{fig:measure_ageing}}
     	\subfloat[]{\includegraphics[width=4cm]{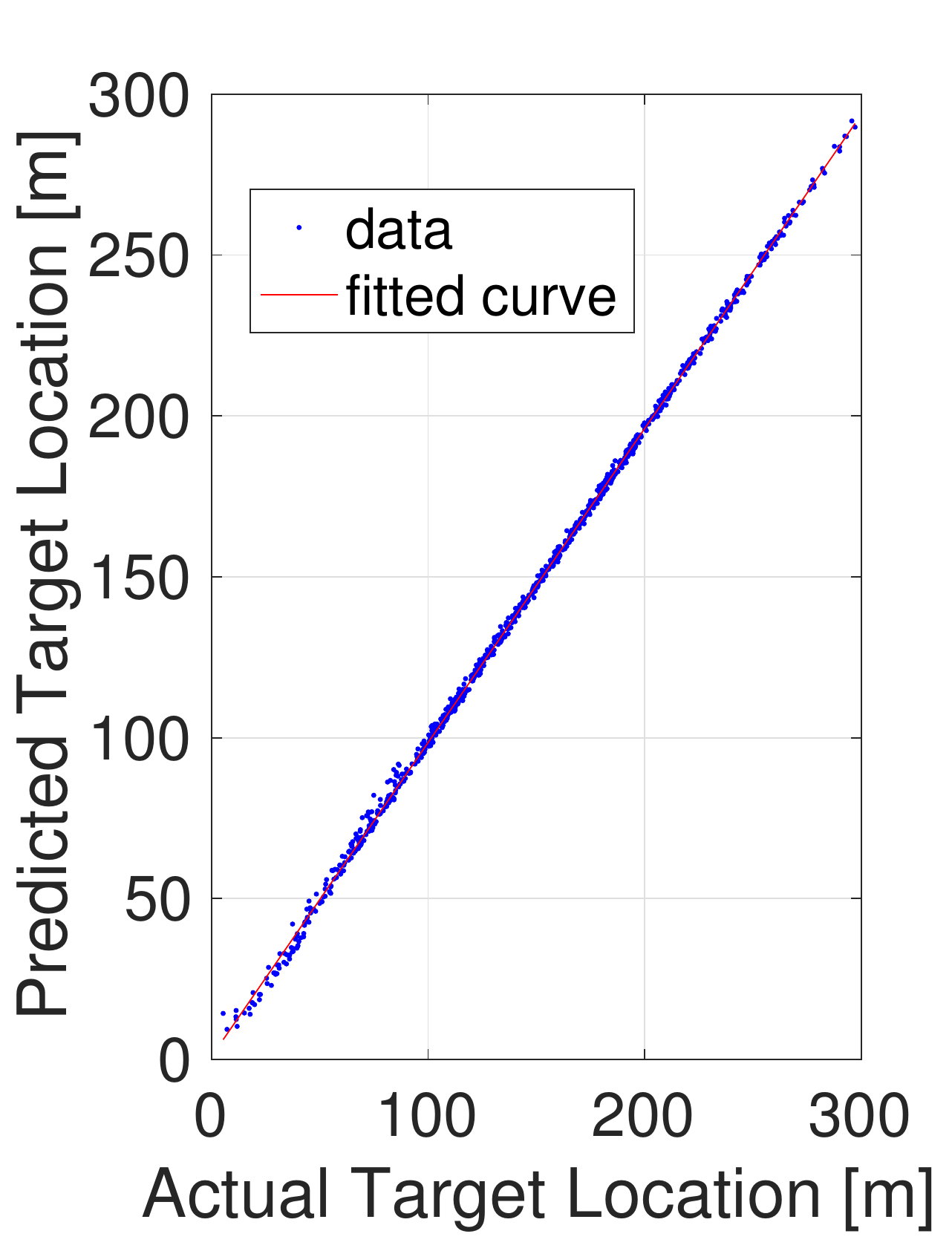} \label{fig:measure_location}}
	 \vspace*{-5pt}
     	\caption{Robustness evaluation results of our proposed scheme with $(n_\mathsf{TR}, n_\mathsf{TE})=(3600, 1000)$ for (a) predicting the equivalent age of a homogeneous degradation, and (b) the target location for an LD.}
	 \label{fig:measure_robustness}
\end{figure}


Although the WT model we use in~\eqref{eqn:WTRegion} is representative of realistic WT degradations, we explored the robustness of our solution to possible behavior deviations that could be seen in practically encountered WT degradations. To this end, we evaluated our solution by training our machine to associate the behavior of WT degradations {with the model of}~\eqref{eqn:WTRegion}, and then testing its performance by applying real-world WT degradation measurements reported in~\cite[Ch. 6]{mugala2005high}. The results of this exercise can be seen in Fig.~\ref{fig:measure_robustness}. For illustration purposes, we present the result of two evaluations we performed, one each with and without an LD. First, we predict the equivalent age of a cable that is subject to homogeneous WT degradation. In comparison to similar results obtained earlier in Fig.~\ref{fig:predict_severity}(a), we observe in Fig.~\ref{fig:measure_robustness} that the prediction performance is noticeably affected, as expected.
However, our machine is robust enough to provide a satisfactory performance with near unity slope of the fitted line and low individual prediction variance. Next, we focus on analyzing an LD, where we predict its target location, as in Fig.~\ref{fig:localization_results}(a). In this case, we notice in Fig.~\ref{fig:measure_robustness}(b) that the prediction results are still accurate, since locating a degradation mainly relies on the {reflection peaks} caused due to a discontinuity. {The prominent peaks and smooth floor of our proposed PLM-JTFDR make detecting and locating such discontinuities fairly robust to the accuracy of the WT model used.}


{The framework presented} in our work also inspires future engineering endeavors, e.g., investigating the extent of the impact of channel estimation errors on the prediction accuracy, and examining the robustness of our solution when trained with one form of degradation and tested on cables subject to multiple types of degradations and various other non-idealities (e.g., bending of cable {or cable splicings}).

\section{Conclusions}
\label{sec:conclusion}
In this paper, we have proposed methods to reuse power line modems available across the smart-grid distribution network to also {monitor cable health conditions}. We propose a machine learning framework to present an automated cable diagnostics procedure. Our solution includes multiple tasks such as cable aging profile classification, cable degradation severity assessment, and precise degradation localization in case of a localized degradation, performed in a sequential manner. Our simulation results show high detection and prediction accuracies for each of these tasks, which can also be attributed to the enhanced features used for machine training that we extract from our newly designed power line modem synthesized joint-time-and-frequency-domain reflectometry technique, as well as to the new insight gained from the reflected channel we use in addition to the end-to-end channel transfer function. The machine uploaded on the power line modems can be updated as often as required with online firmware upgrades. Our proposed technique provides utilities with a low-cost solution that enables them to harness power line modems as not only communication devices, but also as pervasive grid sensors to continuously monitor the status of the cables, and take preemptive measures to avoid cable in-service failures and resultant power outages.

\appendices
\section{PLM-JTFDR}\label{sec:plcJTFDR}
\begin{figure}[!t]
	\centering
	\includegraphics[width=7cm]{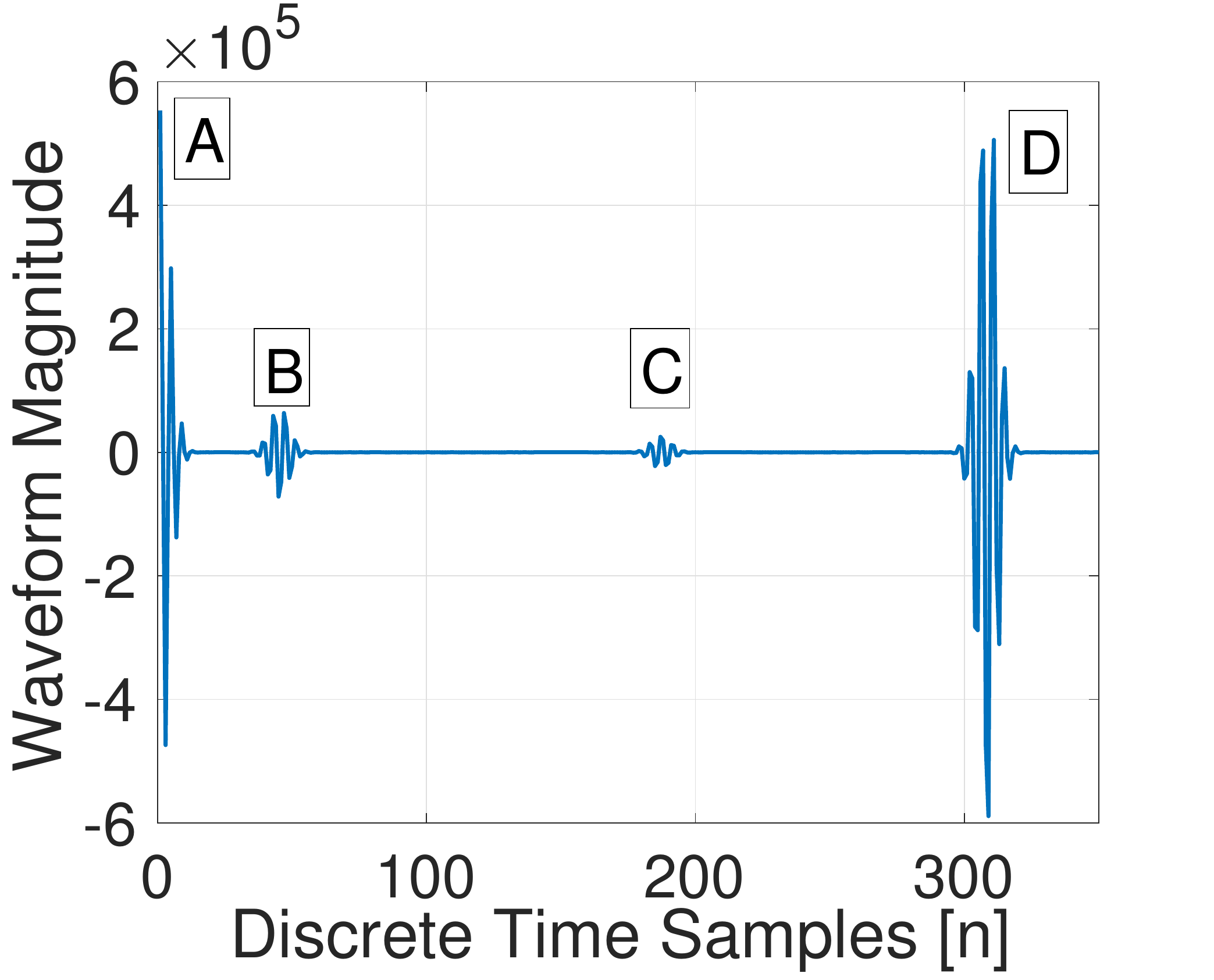}
	\caption{Synthesized JTFDR received signal after auto-correlation.}
	\label{fig:JTFDR_filter}
\end{figure}
\begin{figure}[!t]
	\centering
	\includegraphics[width=7cm]{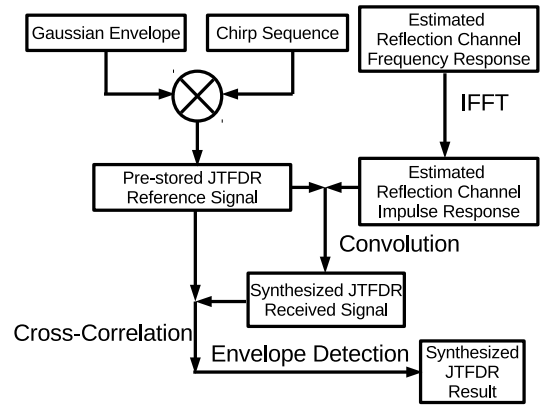}
	\caption{Flowchart of PLM-JTFDR following the underlying JTFDR principle outlined in~\cite{wang2010application}.}
	\label{fig:synthesized_JTFDR}
\end{figure}

Conventional JTFDR involves transmitting a Gaussian enveloped chirp sequence, $s_{\gc}(t) = g(t) \cdot c(t)$, where $g(t)$ and $c(t)$ are {the Gaussian-shaped and the chirp sequences}, respectively, and subsequently sampling and processing the reflected signal, ${\rho}(t)$, to diagnose cable anomalies~\cite{wang2010application}. To emulate this procedure in a PLM, we convolve a pre-generated and pre-stored $s_{\gc}(t)$ with the estimated reflected channel impulse response, ${h}_{\si}  (t)$, to produce the equivalent received JTFDR signal, $\hat{\rho}(t)$, as
\begin{equation}
\hat{\rho}(t) = (s_{\gc} \star {h}_{\si} ) (t), 
\end{equation}
where `$\star$' indicates the linear convolution operation. We then process $\hat{\rho}(t)$ as in conventional JTFDR to obtain locations and magnitudes of peaks in the resultant final waveform, $h_\mathsf{JTFDR}$, for detecting and locating any possible cable degradations. {Specifically}, we compute the cross-correlation signal, $u(t)$, as
\begin{equation}
u(t) = \int_{-\infty}^{\infty} s_{\gc}(\tau) 	\hat{\rho}(t + \tau) \mathrm{d}\tau,
\end{equation}
and the resultant waveform is shown in Fig.~\ref{fig:JTFDR_filter}. To obtain more salient peaks, we compute the envelop of peaks in the waveform of $u(t)$ by passing $|u(t)|$ through a low-pass filter {to obtain $h_\mathsf{JTFDR}$}. We also summarize the operating procedure in Fig.~\ref{fig:synthesized_JTFDR}.

\section{Choice of Number of Training and Testing Samples}
\label{sec:sample_number}

As described in Section~\ref{sec:plcModel}, we use the bottom-up approach to simulate BB-PLC channels subject to {varying degradation and load conditions} as training and testing samples. In such a manner, we independently generate $n_\mathsf{TR}$ and $n_\mathsf{TE}$ number of training and testing samples, respectively. 

We choose $n_\mathsf{TR}$ such that the performance of the machine is saturated given a chosen ML algorithm and a specific set of extracted features, and $n_\mathsf{TE}$ to ensure that we have a sufficient number of testing samples to evaluate the performance of the trained machine to obtain a consistent performance trend in terms of prediction accuracy and prediction variance. Therefore, we choose a large $n_\mathsf{TE}$ for each step of our proposed cable health monitoring scheme. In particular, we choose $n_\mathsf{TE}=1000$ for regression tasks to ensure that a clear trend in the results is observable.

\begin{figure}[!t]
	\centering
	\includegraphics[width=7cm]{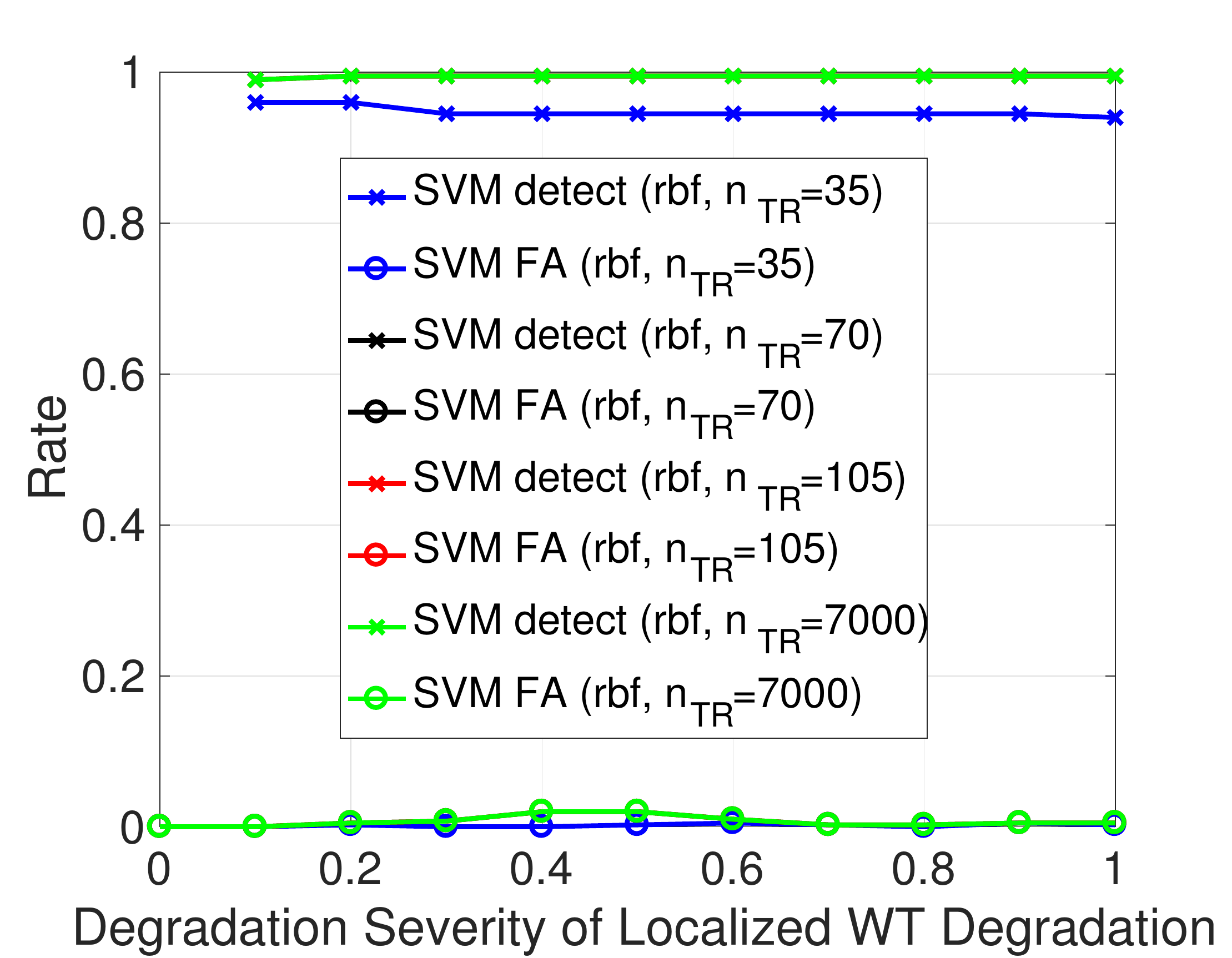}
	\caption{Performance of LD identification versus $n_\mathsf{TR}$.}
	\label{fig:performance_n_tr_classification}
\end{figure}

\begin{figure}[!t]
	\centering
	\includegraphics[width=7cm]{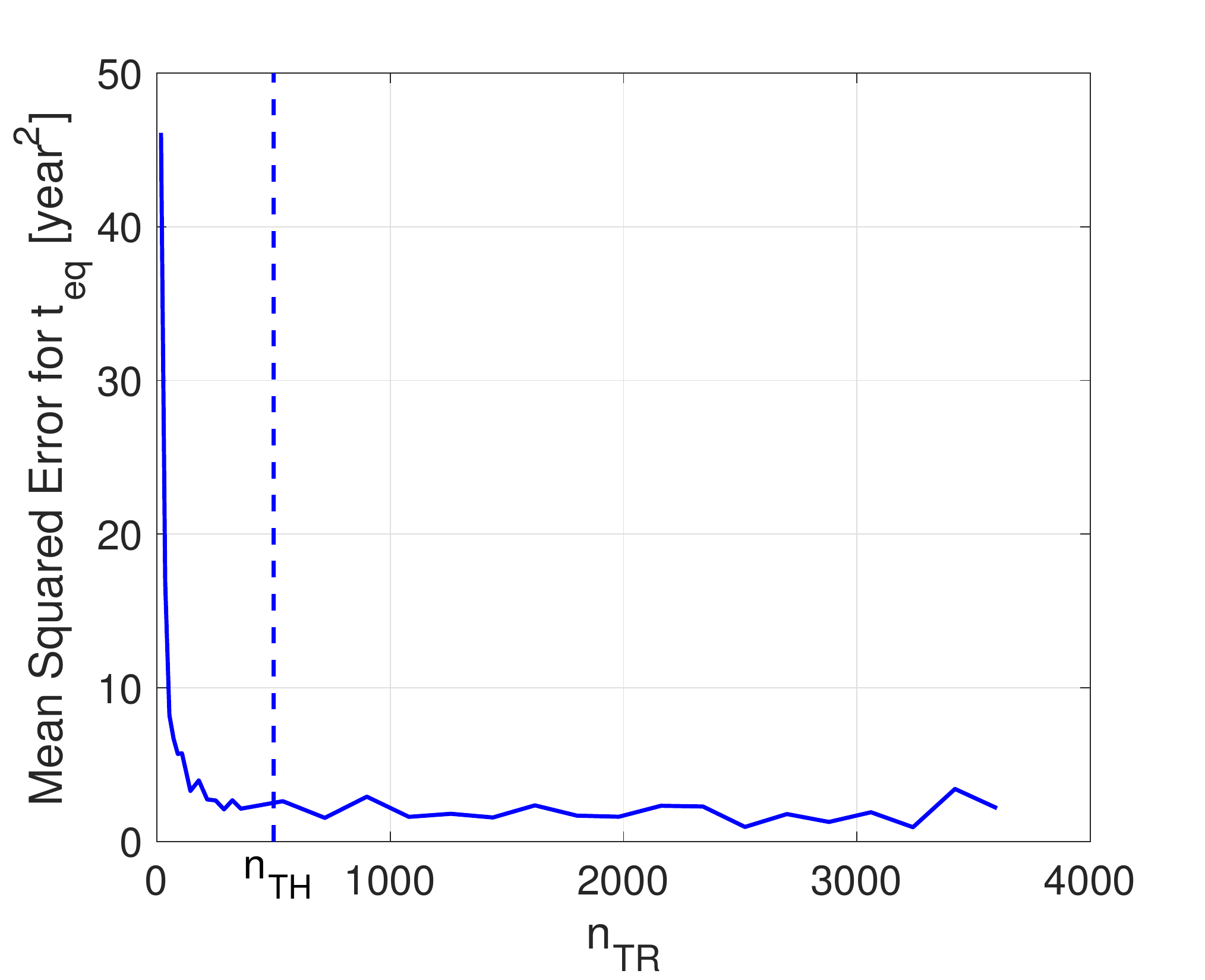}
	\caption{Performance of homogeneous degradation severity assessment versus $n_\mathsf{TR}$.}
	\label{fig:performance_n_tr_ageing}
\end{figure}

\begin{figure}[!t]
	\centering
	\includegraphics[width=7cm]{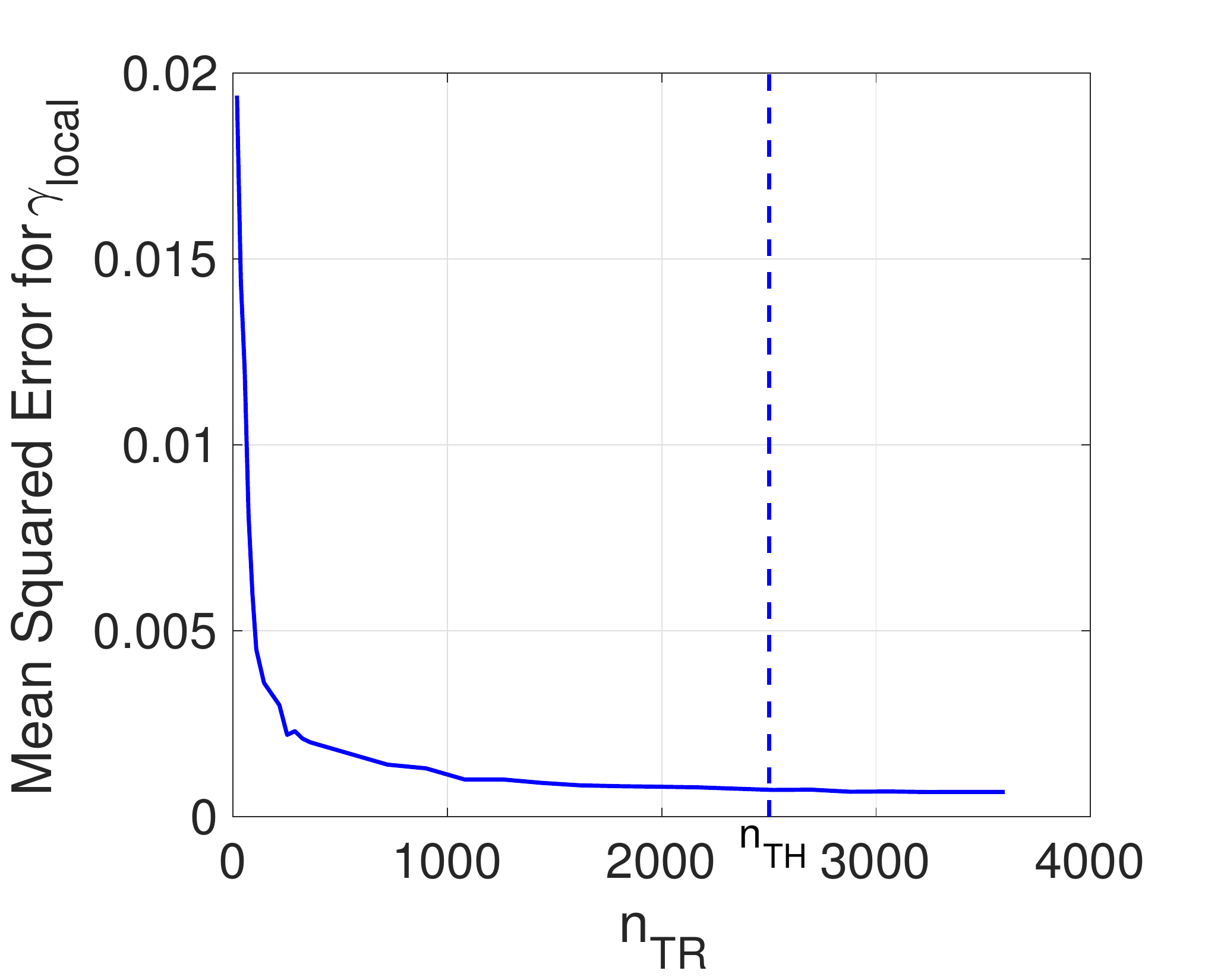}
	\caption{Performance of LD severity assessment versus $n_\mathsf{TR}$.}
	\label{fig:performance_n_tr_gamma}
\end{figure}

However, the choice of $n_\mathsf{TR}$ is more intricate. To this end, we present additional numerical results in Figs.~\ref{fig:performance_n_tr_classification}$-$\ref{fig:performance_n_tr_gamma}, where the variation in performance of the trained machine with varying $n_\mathsf{TR}$ for different ML tasks is shown. The performance of LD identification, and degradation severity assessments, for both a homogeneous aging profile as well as an aging profile with an LD, are shown in Figs.~\ref{fig:performance_n_tr_classification}$-$\ref{fig:performance_n_tr_gamma}, respectively. We notice that while the performance improves with increasing $n_\mathsf{TR}$, it saturates beyond a certain threshold, $n_\mathsf{TH}$. $n_\mathsf{TH}$ is dependent on the ML task, and also on the channel and network conditions. In our investigations, for the tasks in Figs.~\ref{fig:performance_n_tr_classification}$-$\ref{fig:performance_n_tr_gamma}, we have chosen $n_\mathsf{TR}=7000$,  $n_\mathsf{TR}=3600$, and $n_\mathsf{TR}=3600$, respectively, which all well-satisfy $n_\mathsf{TR} > n_\mathsf{TH}$. Specifically, the green curve in Fig.~\ref{fig:performance_n_tr_classification}, and the far-end points in the blue curves of Figs.~\ref{fig:performance_n_tr_ageing}~and~\ref{fig:performance_n_tr_gamma} correspond to the performance shown in Fig.~\ref{fig:branch_location} (with AdaBoost) and Fig.~\ref{fig:predict_severity}. Similarly, for other tasks presented in our work, we choose a sufficiently large $n_\mathsf{TR}$ to ensure a saturated performance of the trained machine.

\section{Strategy to Avoid Over-Fitting}
\label{sec:avoid_overfitting}

The training stage of supervised ML is essentially optimizing the parameters involved in the ML algorithm. When the parameters of the ML model to be trained are large and the number of training samples is low, the parameters are fitted to the small number of training samples and the trained machine lacks the capability of generalization, i.e., introduces over-fitting.  The number of parameters associated with an ML model is closely related to the number of features that we use to train the machine. Typically, ten times the number of features used is the required number of training samples to obtain meaningful results and avoid over-fitting~\cite{mostafa2012learning}. Since our ML implementation uses at most 16 features, we use at least $3600$ training samples ($3600/16=225\gg10$), and therefore avoid over-fitting. We see from Fig.~\ref{fig:performance_n_tr_classification}$-$\ref{fig:performance_n_tr_gamma}, that the performance of the trained machine improves with increasing $n_\mathsf{TR}$ and saturates beyond a threshold, which is smaller than our chosen $n_\mathsf{TR}$.

We note here that we have incurred the difficulty of achieving good results by using a small number of features (at most $16$, in our case) in exchange for eliminating risks of over-fitting. When a small number of features is used, we increase the likelihood of encountering conditions where the trained model is not sufficiently comprehensive to predict the label, i.e., we face the issue of under-fitting. In particular, the SVM enforces sparsity on the number of trained support vectors, which determine the boundary for classification or regression. This avoids over-fitting but could result in under-fitting. For some of the tasks, e.g., equivalent age prediction, using an SVM does result in under-fitting. Therefore, we use the boosting technique in such cases, which is able to avoid under-fitting and train a machine with sufficient comprehensiveness to predict the label. In this context, we {also note} that we use the boosting algorithm for a large portion of our tasks, {which is well known to be robust to over-fitting~\cite{freund1999short}.} In such a case, we can simply {tune} its iteration limit to control the complexity of the trained model and obtain a trained machine that is not subject to over-fitting.

\bibliographystyle{ieeetran}
\bibliography{TestBibTex}{}

\end{document}